\documentclass{article}
\usepackage[a4paper, margin=1in]{geometry}

\usepackage[table]{xcolor}
\usepackage[hyphens]{url}
\usepackage{booktabs}
\usepackage{graphicx}
\usepackage{todonotes}
\usepackage[bookmarks=true]{hyperref} % draft for non-clickable links
\usepackage{diagbox}
\usepackage{rotating}
\usepackage{multirow}
\usepackage{tabularx}
\usepackage{authblk}

\newcommand\clearrow{\global\let\rowmac\relax}
\clearrow
\PassOptionsToPackage{table, dvipsnames}{xcolor}

\newcommand{\specialcell}[2][c]{%
	\begin{tabular}[#1]{@{}l@{}}#2\end{tabular}}

\newcommand{\myparagraph}[1]{\vspace{0.1cm}\noindent{\it \textbf{#1}.}}

\title{Intercepting Hail Hydra: Real-Time Detection of Algorithmically Generated Domains}

\author[1]{Fran Casino}
\author[1]{Nikolaos Lykousas}
\author[2]{Ivan Homoliak}
\author[1,3]{Constantinos Patsakis}
\author[4]{Julio Hernandez-Castro}
\affil[1]{Department of Informatics, University of Piraeus, Greece}
\affil[2]{Faculty of Information Technology, Brno University of Technology, Czech Republic}
\affil[3]{Information Management Systems Institute of Athena Research Center, Greece}
\affil[4]{School of Computing, University of Kent, United Kingdom.}
\date{}

\providecommand{\keywords}[1]{\textbf{\textit{Index terms---}} #1}

\usepackage{graphicx}

\begin{document}

\maketitle

\begin{abstract}

A crucial technical challenge for cybercriminals is to keep control over the potentially millions of infected devices that build up their botnets, without compromising the robustness of their attacks.
A single, fixed C\&C server, for example, can be trivially detected either by binary or traffic analysis and immediately sink-holed or taken-down by security researchers or law enforcement.
Botnets often use Domain Generation Algorithms (DGAs), primarily to evade take-down attempts. DGAs can enlarge the lifespan of a malware campaign, thus potentially enhancing its profitability. They can also contribute to hindering attack accountability.

In this work, we introduce \texttt{HYDRAS}, the most comprehensive and representative dataset of Algorithmically-Generated Domains (AGD) available to date. The dataset contains more than 100 DGA families, including both real-world and adversarially designed ones. We analyse the dataset and discuss the possibility of differentiating between benign requests (to real domains) and malicious ones (to AGDs) in real-time. The simultaneous study of so many families and variants introduces several challenges;
nonetheless, it alleviates biases found in previous literature employing small datasets which are frequently overfitted, exploiting characteristic features of particular families that do not generalise well.
We thoroughly compare our approach with the current state-of-the-art and highlight some methodological shortcomings in the actual state of practice. The outcomes obtained show that our proposed approach significantly outperforms the current state-of-the-art in terms of both classification performance and efficiency.

\end{abstract}
\keywords{
Malware, Domain Generation Algorithms, Botnets, DNS, Algorithmically Generated Domain}

\section{Introduction}

The continuous arms race between malware authors and security researchers has pushed modern malware to evolve into highly sophisticated software, capable of infecting millions of devices. The vast amount of sensitive information that can be extracted from compromised devices, coupled with the harnessing of their resources and processing power, provides a wide range of monetisation methods fuelling a flourishing worldwide underground economy.

\begin{figure}[t]
    \vspace{-0.6cm}
    \centering
    \includegraphics[width=0.9\columnwidth]{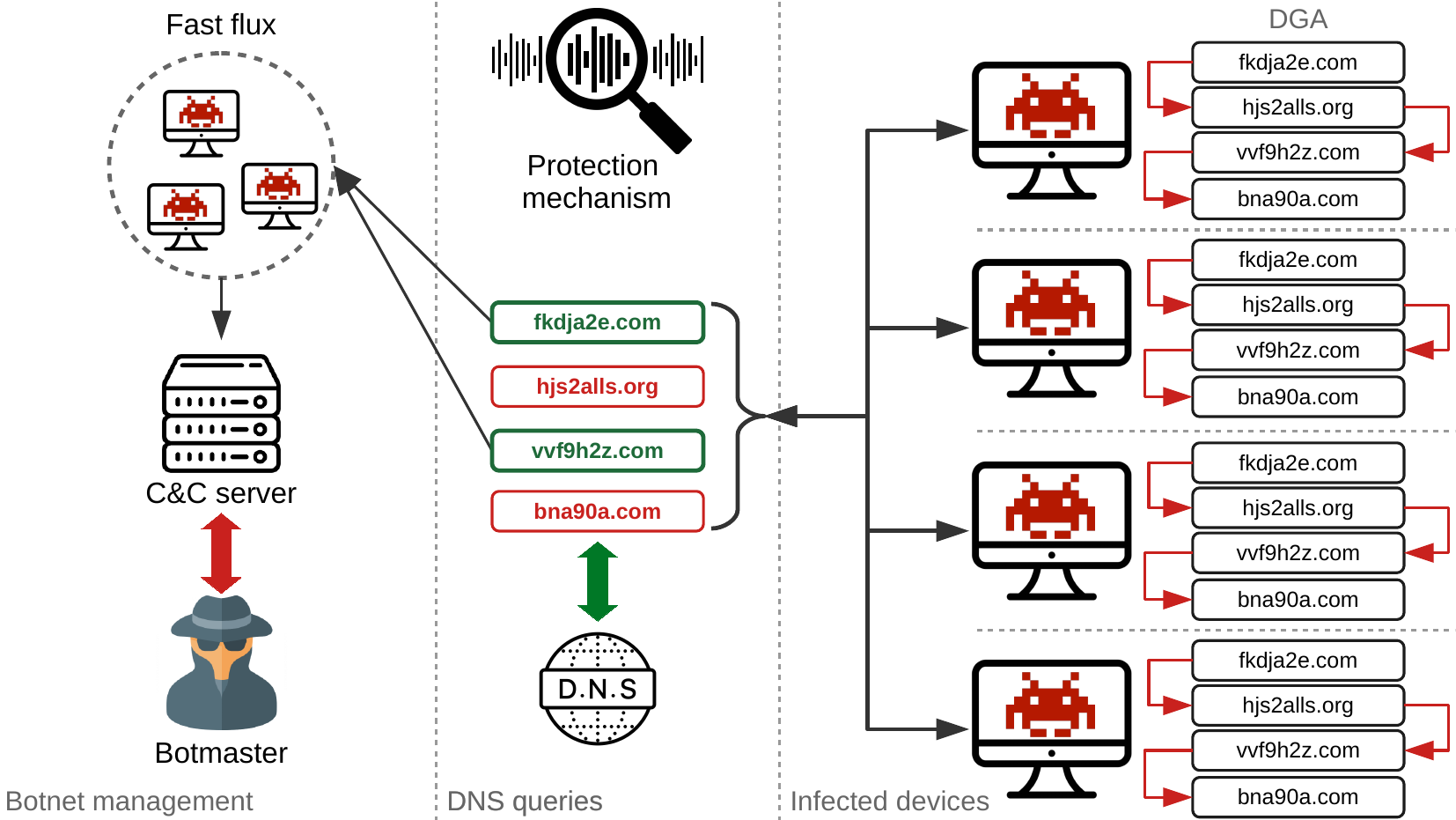}
    \caption{The \textit{modus operandi} of a typical DGA-powered botnet~\cite{PATSAKIS2020101614}. }
    \label{fig:dga_use_case}
    \vspace{-0.4cm}
\end{figure}

While device infection is the key that paves the way in, the main objectives are generally persistence and orchestration.
An orchestrating entity, the botmaster, manages infected devices (bots) which in many cases can scale to the order of millions, creating a botnet~\cite{SINGH201928}.
The botmaster manages a Command and Control (C\&C) server that communicates with the bots. This communication must preserve some degree of unlinkability to thwart any attempts to identify the botmaster.
To ensure unlinkability, and as a counter-measure against take-down operations, botnets frequently make use of domain fluxing~\cite{6175908,8532279} through Domain Generation Algorithms (DGAs). DGAs produce a vast amount of domain names, which bots try to communicate with iteratively to find the actual C\&C server. However, only a small part of them is registered and active, creating a hydra effect~\cite{nadji2013beheading}.
The botmaster may regularly pivot control between domains, thus hampering the task of seizing control of the botnet. This is helped by the fact that an outsider cannot determine which domains will be used, nor statically block all these requests.
The latter stems from the fact that there are too many domains, and the seed yielding a particular sequence of domains might be unknown or change frequently.
\autoref{fig:dga_use_case} illustrates the \textit{modus operandi} of a typical DGA-powered botnet.

Currently, there are several families of DGAs employed by various malware with varying rates of requests and different characteristics.
This heterogeneous landscape hinders the timely and accurate detection of an Algorithmically-Generated Domain (AGD)~\cite{yadav2012} request, which could serve as a precise indicator of compromise (IoC) of a host at the network level.
Recent research tries to categorise DNS requests per DGA, often exploiting WHOIS-based features.
From the perspective of an ISP, CSIRT or CERT such an approach might be beneficial.
However, we argue that in terms of endpoint security that strategy cannot be considered adequate. First, the network operator does not generally know which concrete DGA is utilised by the malware that infected a given host in her network. Second, the utilisation of the WHOIS database introduces a significant time delay that in many situations cannot be tolerated.

\subsection{Contributions}
Motivated by the continuous evolution of DGAs we introduce a dataset collecting real-world domains called \texttt{HYDRAS}, which consists of more than 95 million domains belonging to 105 unique DGA families. To the best of our knowledge, this is the largest and most representative DGA dataset to date.

During the analysis of our dataset, the possibility of differentiating between benign and malicious requests in real time is discussed, as well as the identification of the malware families using them.
Based on information learned from the analysis of our dataset, a novel feature set is designed and implemented, which includes lexical and statistical features over the collected DGAs, as well as English gibberish detectors. %IH: methodology is not correct to me; it should be the approach
Using the proposed feature set and a Random Forest as a representative of ensemble classifiers, we perform a thorough evaluation of our dataset and show that our feature set together with the Random Forest classifier outperform the state-of-the-art approaches in terms of both classification performance and overhead.

Next, inherent biases in related works are highlighted. These biases can be attributed to the suboptimal selection of datasets and/or features, preventing their application in general, real-world scenarios.
For example, employing a dataset comprising only a few families that exhibit very characteristic patterns might ease the classification task, providing accurate detection rates (e.g., the generators for \texttt{cryptolocker}, \texttt{ramnit},  \texttt{geodo}, \texttt{locky}, \texttt{tempedreve}, \texttt{hesperbot}, \texttt{fobber} and \texttt{dircrypt} provide a uniform distribution of letters \cite{18,2}), but will inevitably lead to ad-hoc solutions that are too specific and cannot be generalised.
This is, unfortunately, a common practice in the existing literature.

A typical example is only considering families like \texttt{bamital}, \texttt{CCleaner} or \texttt{chir} in the datasets, which all produce hexadecimal values of specific length as second-level domains (SLDs).
It is obvious that one can easily differentiate benign domains from such DGAs with almost 100\% accuracy by merely checking whether the SLD is a hex value of a specific length.
Nonetheless, not all DGAs families are so easy to detect in real scenarios.

Due to the particularities of this research field and the methodologies used by the current state-of-the-art, we also highlight some recommendations for fairer future evaluations. These are particularly relevant for comparing the results of our experiments with other approaches.

\subsection{Organization}
The rest of this work is organised as follows.
In \autoref{sec:background}, DGA-related preliminaries are briefly described, and in \autoref{sec:related-work} a thorough review of related work is provided.
Then, in \autoref{sec:data}, our dataset is detailed.  Afterwards, we describe our approach, including methodology, feature extraction and the tools and algorithms employed.
In \autoref{sec:proposed-features}, we describe the proposed features.
We provide the results of our experiments in \autoref{sec:classification}, \autoref{sec:adversarial} and \autoref{sec:performance}, where they are compared to the state-of-the-art.
In \autoref{sec:discussion}, an analysis of the outcomes as well as a methodological analysis and comparison with the literature is provided.
Finally, the paper concludes in \autoref{sec:conclusions}, discussing open issues for future research.

\section{Background}
\label{sec:background}

\begin{figure}[t]
     \vspace{-0.3cm}
    \centering
    \includegraphics[trim={0 0 23cm 0},clip,width=.5\columnwidth]{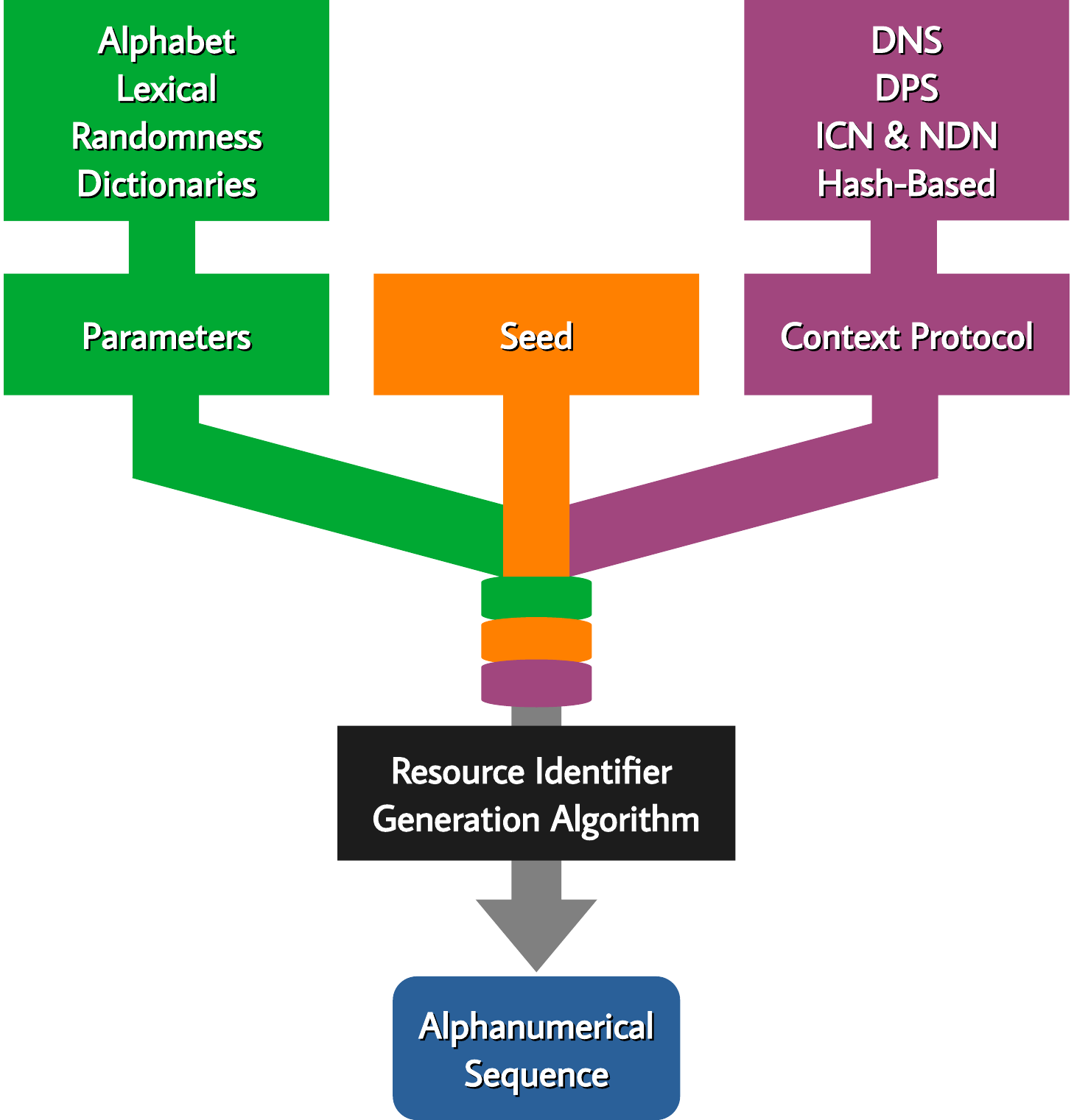}
    \caption{RIGA generation flow, including diverse context protocols (e.g., DPS and DNS).}
    \label{fig:riga}
    \vspace{-0.4cm}
\end{figure}

DGAs are one of the main pillars behind the success of botnets.
They were first conceived more than a decade ago, and they have been steadily refined over the years by successive generations of malware developers.
These algorithms generate a set of AGDs to communicate with C\&C servers, thus eliminating the risks associated with using static IP addresses~\cite{203628,nadji2017still}.
In~\cite{Patsakis2019}, the authors generalise the notion of DGAs by extending them to other protocols beyond DNS, and they propose the term Resource Identifier Generation Algorithms (RIGAs).
The authors show how decentralised permanent storage (DPS) has some potential drawbacks and exploitable characteristics for armouring a botnet, a fact that has already been exploited in the real world~\cite{ipfsstorm} due to the immutability properties of DPS.
\autoref{fig:riga} depicts the hierarchy of RIGAs.

In its most basic form, DGAs create a set of domain names by using a deterministic pseudo-random generator (PRNG)~\cite{7535098,6175908}. Therefore, (infected) devices belonging to a botnet query a set of domains generated by the DGA until they are correctly resolved to a valid IP, corresponding to the C\&C server.
Since the location of the C\&C server dynamically changes, blacklisting domains is a very inefficient protection technique.
Additionally, this makes seizing the botnet much more difficult, since one would need to take (register) all domain names generated by the DGA (with a given seed) for disrupting the botmaster only for a short amount of time. This process will generally be very costly, typically involving thousands of domain names for stopping the botnet for just a day.
Hence, the botmaster benefits from this asymmetry between the high ratio of generated domains to registered ones. That makes her operation cheap, as compared to the cost of defending against it, which involves registering all possible domains.

\section{Related Work}\label{sec:related-work}
\subsection{Traditional Approaches}
% \cite{Bharathi2019678,attardi2018bidirectional,chen2018towards,yu2017inline,18}.
% counter DGAs

According to the literature, there are two main DGA families:
(i) \textbf{Ran\-dom-based} DGA methods, which use a PRNG to generate a set of characters that form a domain name,
and (ii) \textbf{Dictionary/Wordlist-based} DGA methods, which use a dictionary to produce domains.
Nevertheless, one may also consider other types of DGA families, which use more subtle approaches, i.e. valid domains that were previously hacked to hide their C\&C servers (i.e. domain shadowing)~\cite{Liu2017} as well as DGAs that generate domains that are very similar to existing valid domains~\cite{johannesbader}, further hindering the detection task.
Considering the dependency of the pre-shared secret (or seed) on time, Plohmann et al.~\cite{197187} further categorise DGAs into: (i) \textit{time-independent and deterministic}, (ii) \textit{time-dependent and deterministic}, and (iii) \textit{time-dependent and non-deterministic}.

In the case of random-based DGA detection, a common practice is to analyse some features of the domain names and their lexical characteristics to determine whether a DGA has generated them \cite{Aviv2011,6151233}. Moreover, auxiliary information such as WHOIS and DNS traffic (e.g. frequent NXDomain responses) is often used to detect abnormal behaviours \cite{Zhou2013DGABasedBD,5762763,1}. Other approaches use machine learning-based techniques and combine the previous information to identify Random-based DGA such as in \cite{5762763,yadav2012,yadavgraph,7163279}.

Due to their inherent construction, wordlist-based DGA detection represents a challenging task for classifiers. In this regard, the common approach is to use machine learning approaches (e.g., feature-based classification and deep learning) to distinguish between benign and malicious DGAs. The use of random forest classifiers (RF) based on a set of features such as word correlation, frequency, and part-of-speech tags was first proposed in~\cite{yang2018novel}. Similarly, Selvi et al. suggested the use of RF with masked n-grams as a feature, achieving a remarkable accuracy in the binary classification task \cite{SELVI2019156}.
Berman~\cite{berman2019dga} put forward a methodology based on Capsule Networks (CapsNet) to detect AGDs; the author compared his method with well-known approaches such as RNNs and CNNs, and the outcomes showed similar accuracy yet better computational cost. Xu et al.~\cite{XU201977} suggested the combination of n-gram and a deep CNNs to create an n-gram combined with character-based domain classification (n-CBDC) model that does not require domain feature extraction.
Vinayakumar et al.~\cite{Vinayakumar2019} implemented a set of deep learning architectures with Keras and classical machine learning algorithms to classify DGA families.
Their best configuration uses RNNs and SVMs with a radial basis function (SVM-RBF).
Yang et al.~\cite{9072447} present a heterogeneous deep neural network framework, which extracts the local features of a domain name as well as a self-attention based Bi-LSTM to extract further global features.
Their outcomes showed higher accuracy than traditional DGA classifiers. Finally, a recent approach based on the probabilistic nature of wordlist-based DGAs was proposed in \cite{PATSAKIS2021102725}. In their work, Patsakis et al., proposed the combination of feature-based extraction with a probabilistic-based threshold to fully capture wordlist-based AGDs. Moreover, their method was capable of detecting both real-world and custom DGAs created to fool traditional detectors.

%%%%%%%%%%%

%%%%%%%%%%%%%

\subsection{Adversarial and Anti-Forensic Approaches}
Recently the exciting development of deploying anti-forensic techniques in DGAs has become popular. This aims to create hard-to-detect DGA families and to fight against high performing classifiers. % while considering more sophisticated attack scenarios.
Anderson et al.~\cite{Anderson2016} proposed a generative adversarial network (GAN), which can learn from and bypass classical detectors.
Afterwards, they improved the performance of AGD detectors after training them with the data generated by the GAN.
Alaeiyan et al.~\cite{ALAEIYAN2020} proposed a DGA family created with a genetic algorithm considering lexical features such as pronounceability.
Their experiments showed that such a DGA family was hard to detect by classical approaches.
In a similar vein, Yun et al.~\cite{8936543} used n-gram distribution and the pronounceability/readability of domains as a basis to create a novel DGA based on neural language models and the Wasserstein GAN (WGAN), which reduced detection rates in traditional DGA techniques.

Spooren et al.~\cite{Spooren2019} showed that their deep learning RNN performs significantly better than classical machine learning approaches.
Besides, the authors stressed that one of the issues of manual feature engineering is that an adversary may adapt her strategy if she knows which features were used in the detection. % and proposed a set of properly crafted DGAs that bypass these classifiers.
Fu et al.~\cite{7852496} proposed two DGAs using hidden Markov models (HMMs) and probabilistic context-free grammars, which were tested on state-of-the-art detection systems.
Their results revealed their DGAs hindered the detection rate known detectors.
%Patsakis and Casino~\cite{patsakis2019exploiting} proposed a probabilistic method to detect wordlist-based DGAs, which exploited the fact that some DGAs use a relatively limited dictionary, often resulting in word repetitions and the ``birthday problem''. % to detect the domain queries over a threshold.

Finally, and due to the widespread use of covert or encrypted communication channels in DNS (e.g., DNSCurve, DNS over HTTPS and DNS over TLS) and in C\&C connections in general~\cite{zander2007survey,homoliak2014nba}, malware creators have an additional tool to hide their activity, rendering many traditional DGA detection useless.

Nevertheless, as shown by Patsakis et al.~\cite{PATSAKIS2020101614}, NXDomain detection can still be carried out in such a scenario. This also applies to feature extraction, so DGA families can still be classified with high performance.

\section{The HYDRAS Dataset}
\label{sec:data}
In this section, the \texttt{HYDRAS} dataset is introduced, which consists of a collection of benign and AGD domains, both real-world as well as adversarial.  The name of the dataset originates from the insightful parallelism suggested by Nadji et al. \cite{nadji2013beheading} between DGA-powered botnets and the mythical ancient Greek monster.

Benign domains are sampled from the Alexa 1M dataset.
But since the Alexa dataset contains sites, not domains, it had to preprocessed.
First, all top-level domain names (e.g., \textit{.com}, \textit{.org}) are removed from each entry and only the SLD are kept.
Then, the duplicates were pruned since some web pages have multiple entries in the dataset (e.g., \textit{google.com} and \textit{google.co.in}) or been subdomains of identical services (e.g., various blogs of \textit{blogspot.com}).
Finally, all internationalised domain names were removed, since they are encoded using Punycode\footnote{https://tools.ietf.org/html/rfc3492} representation.
After preprocessing the 1M Alexa dataset, the final dataset contains 915,994 unique domains.

\begin{table}[t]
	\centering
	\caption{Distribution of records per DGA in our dataset. DGAs in {\color{teal}green} denote those which were frequently underrepresented, so they were run to create more samples, while {\color{purple} purple} indicates adversarial ones.}
	\label{tab:sample_stats}
	\resizebox{\textwidth}{!}{%
	\begin{tabular}{>{\bfseries}lr>{\bfseries}lr>{\bfseries}lr>{\bfseries}lr}
		\toprule
		Class                & \textbf{Support} & Class              & \textbf{Support}  & Class                  & \textbf{Support} & Class         & \textbf{Support}  \\
		\midrule
bamital	&	86892	&	feodo	&	247	&	omexo	&	41	&	sisron	&	2580	\\
banjori	&	439423	&	fobber	&	2000	&	padcrypt	&	246096	&	sphinx	&	174726	\\
bedep	&	7814	&	fobber\_v1	&	298	&	pandabanker	&	32484	&	{\color{teal}suppobox}	&	98304	\\
beebone	&	72	&	fobber\_v2	&	299	&	pitou	&	74314	&	sutra	&	3295	\\
bigviktor	&	999	&	gameover	&	22723000	&	{\color{teal}pizd}	&	16384	&	symmi	&	65	\\
blackhole	&	732	&	geodo	&	576	&	post	&	66000	&	szribi	&	20661	\\
{\color{teal}bobax/kraken/oderoor}	&	30459	&	{\color{teal}gozi}	&	163529	&	proslikefan	&	218399	&	{\color{teal}tempedreve}	&	13323	\\
{\color{teal}ccleaner}	&	12000	&	goznym	&	364	&	pushdo	&	380427	&	tinba	&	72719	\\
chinad	&	750312	&	gspy	&	100	&	pushdotid	&	6000	&	tinynuke	&	52832	\\
chir	&	100	&	{\color{teal}hesperbot}	&	16512	&	pykspa	&	1996763	&	tofsee	&	2100	\\
conficker	&	2082010	&	infy	&	5220	&	pykspa\_v1	&	44688	&	torpig	&	18716	\\
corebot	&	20931	&	{\color{purple}khaos}	&	10000	&	pykspa\_v2\_fake	&	798	&	tsifiri	&	59	\\
cryptolocker	&	368196	&	{\color{teal}kingminer}	&	252	&	pykspa\_v2\_real	&	198	&	ud2	&	491	\\
cryptowall	&	56624	&	locky	&	994381	&	pykspa2	&	1248	&	ud3	&	20	\\
darkshell	&	49	&	{\color{teal}madmax}	&	4850	&	pykspa2s	&	9960	&	ud4	&	70	\\
{\color{purple}deception}	&	149854	&	makloader	&	256	&	qadars	&	630127	&	{\color{teal}vawtrak}	&	17807	\\
{\color{purple}deception2}	&	149908	&	matsnu	&	40050	&	qakbot	&	4579999	&	{\color{teal}vidro}	&	62567	\\
diamondfox	&	279	&	{\color{teal}mirai}	&	2716	&	qhost	&	23	&	vidrotid	&	101	\\
{\color{teal}dircrypt}	&	11210	&	modpack	&	107	&	qsnatch	&	1246482	&	virut	&	23669176	\\
dmsniff	&	70	&	monerodownloader	&	2995	&	ramdo	&	6000	&	volatilecedar	&	498	\\
dnschanger	&	1499578	&	monerominer	&	364271	&	ramnit	&	150662	&	wd	&	32172	\\
{\color{teal}dromedan}	&	10000	&	murofet	&	13824213	&	ranbyus	&	578080	&	{\color{teal}xshellghost}	&	12001	\\
dyre	&	2046998	&	murofetweekly	&	600000	&	{\color{teal}redyms}	&	91	&	xxhex	&	1900	\\
ebury	&	2000	&	mydoom	&	2599	&	rovnix	&	207996	&	{\color{teal}zloader}	&	29992	\\
ekforward	&	3649	&	necurs	&	12751075	&	{\color{teal}shifu}	&	2554	& & \\
emotet	&	431048	&	nymaim	&	700102	&	shiotob/urlzone/bebloh	&	37031	&  \textbf{Total}& \textbf{95,325,598}\\
enviserv	&	500	&	{\color{teal}nymaim2}	&	110511	&	{\color{teal}simda}	&	24345	& & \\
		\bottomrule
	\end{tabular}
	}
\end{table}
The use of small and unrepresentative datasets, unfortunately very frequent in the literature, leads to several biases and other issues that can easily lead towards wrong analysis and misleading conclusions.
For instance, the public feed of DGAs provided by the Network Security Research Lab at 360\footnote{\url{https://data.netlab.360.com/dga/}} as well as the DGArchive \cite{197187} provide real-world datasets with millions of samples from many DGA families.
Nonetheless, despite the numerous samples in both these datasets, many malware families are significantly underrepresented.
A demonstrative example is the \texttt{xshellghost} family in the 360 dataset, which contains only a single sample at the time of writing. Thankfully, the researchers at 360 have reversed the code of this DGA\footnote{\url{https://github.com/360netlab/DGA/blob/master/code/xshellghost/dga.py}}.

Since the provision of many samples is required to perform an adequate evaluation of any detection technique, we utilised the available code of poorly represented DGAs to enlarge our dataset.
The dataset was initialised with several public DGA repositories, e.g., J. Bader's~\cite{badergithub}, A. Abakumov's \cite{abakumovgithub}, and P. Chaignon's \cite{chaignongithub}. As explained above,  we additionally used DGA code available at these and other repositories to generate additional samples for underrepresented DGA families. In these cases, a few random seeds and/or an extended date range to obtain new samples was used.

Since we used the code of the DGAs, the added domains have identical characteristics to original ones and might occur in the real-world. Thus, these AGDs could have been collected in a real setting.
Moreover, the SLDs of three adversarially designed DGAs, namely \texttt{deception}, \texttt{deception2} \cite{Spooren2019} and \texttt{khaos} \cite{8936543} were added.

In summary, our dataset consists of 95,325,598 AGDs belonging to a total of 109 families, from which 105 are unique.\footnote{A few DGAs are used by multiple families.}
The families included, along with their corresponding number of collected samples, are reported in \autoref{tab:sample_stats}.
The dataset is available for download at \url{https://zenodo.org/record/3965397}~\cite{hydra_dataset}.

\section{Proposed Features}
\label{sec:proposed-features}
We thoroughly analysed the AGDs in our dataset, as well as the ideas behind existing AGD detection approaches in the literature.
We found out that the basic strategy for detecting non-wordlist-based DGAs is to take advantage of the fact that they, in general, make little effort to be human-memorable, as they typically are randomly generated.
Moreover, even if they show a high correlation with readable words in terms of vowel/consonant usage, etc., the generated domains are expected to contain zero to only a few words having a short length.

\subsection{Approach to Feature Extraction}
A general description of our approach is as follows: On receiving a domain name, we first cache it to see correlations with previous ones. Then, we try to determine whether the SLD matches some specific patterns, e.g., whether it is a hex value, its combination of vowels/consonants, length, etc. Later, after removing all digits, we try to break the remaining characters into words.
Within these words, the short ones (e.g., stop words, articles) are pruned and study the remaining to determine whether they are real words or just gibberish. Moreover, the entropy of the domain is computed and a subset of the patterns created during the correlation process.
All the above provide us with several features that can be efficiently used to determine whether a domain name is benign or not, without the need for external information (e.g., WHOIS) or waiting for the DNS resolution revealing whether it is an NXDomain. In this way, a significant number of requests are pruned, regardless of their outcome.

\begingroup
\medmuskip=0mu
\thickmuskip=0mu
\begin{table}[!ht]
\vspace{-0.3cm}
\centering
\caption{Features used in our approach and their corresponding description.}
\label{tab:parameters}
%\rowcolors{2}{gray!25}{white}
\tiny
%\resizebox{\textwidth}{!}{%
\begin{tabular}{>{\bfseries}p{1.2in}p{1in}p{2.5in}}
\toprule
\textbf{Feature Set} & \textbf{Notation} & \textbf{Description} \\
\midrule
\multirow{7}{*}{Alphanumeric Sequences} & $Dom$ & Domain without TLD   \\   % txt1
&$Dom-D$  & $Dom$ without digits  \\   % txt3
&$Dom-3G$  & Set of 3-grams of $Dom$  \\
&$Dom-4G$  & Set of 4-grams of $Dom$     \\
&$Dom-5G$  & Set of 5-grams of $Dom$   \\   %
&$Dom-W$ & Domain concatenated words  \\
&$Dom-WS$ & Domain concatenated words with spaces  \\  %txt2
&$Dom-WD$ & $Dom-D$ concatenated words \\
&$Dom-WDS$ & $Dom-D$ concatenated words with spaces  \\% txt4
&$Dom-W2$ & Domain concatenated words of length $>$ 2  \\  % txt5
&$Dom-W3$ & Domain concatenated words of length $>$ 3  \\   % txt6
\midrule
\multirow{5}{*}{Statistical Attributes}& $L-HEX$ & The domain name is represented with hexadecimal characters \\
& $L-LEN$ & The length of $Dom$  \\
& $L-DIG$ & The number of digits in $Dom$   \\
& $L-DOT$ & The number of dots in the raw domain \\
& $L-CON-MAX$ & The maximum number of consecutive consonants $Dom$ \\
& $L-VOW-MAX$ & The maximum number of consecutive vowels $Dom$ \\
& $L-W2$ & Number of words with more than 2 characters in $Dom$ \\
& $L-W3$ & Number of words with more than 3 characters in $Dom$ \\ %words longer than 3 chars in txt1
 \midrule
 \multirow{8}{*}{Ratios} & $R-CON-VOW$ & Ratio of consonants and vowels of  $Dom$ \\
&$R-Dom-3G$  & Ratio of benign grams in $Dom-3G$ \\   %
&$R-Dom-4G$  & Ratio of benign grams in $Dom-4G$    \\  %
&$R-Dom-5G$  & Ratio of benign grams in $Dom-5G$  \\   %
& $R-VOW-3G$ & Ratio of grams that contain a vowel in $Dom-3G$ \\
& $R-VOW-4G$ & Ratio of grams that contain a vowel in $Dom-4G$ \\
& $R-VOW-5G$ & Ratio of grams that contain a vowel in $Dom-5G$ \\
&$R-WS-LEN$ & $Dom-WS$ divided by $L-LEN$ \\
&$R-WD-LEN$ & $Dom-WD$ divided by $L-LEN$ \\
&$R-WDS-LEN$ & $Dom-WDS$ divided by $L-LEN$  \\
&$R-W2-LEN$ & $Dom-W2$ divided by $L-LEN$ \\
&$R-W2-LEN-D$ &  $Dom-W2$ divided by $Dom-D$ \\
&$R-W3-LEN$ & $Dom-W3$ divided by $L-LEN$ \\
&$R-W3-LEN-D$ &  $Dom-W3$ divided by $Dom-D$ \\

\midrule
\multirow{10}{*}{Gibberish Probabilities}&$GIB-1-Dom$  & Gibberish detector 1 applied to $Dom$ \\
&$GIB-1-Dom-WS$  & Gibberish detector 1 applied to $Dom-WS$ \\% GIB1 the higher the better
&$GIB-1-Dom-D$  & Gibberish detector 1 applied to $Dom-D$ \\
&$GIB-1-Dom-WDS$  & Gibberish detector 1 applied to $Dom-WDS$ \\
&$GIB-1-Dom-W2$  & Gibberish detector 1 applied to $Dom-W2$ \\
&$GIB-1-Dom-W3$  & Gibberish detector 1 applied to $Dom-W3$ \\
&$GIB-2-Dom$  & Gibberish detector 2 applied to $Dom$ \\    % gib2 the higher the worse
&$GIB-2-Dom-WS$  & Gibberish detector 2 applied to $Dom-WS$  \\
&$GIB-2-Dom-D$  & Gibberish detector 2 applied to $Dom-D$ \\
&$GIB-2-Dom-WDS$  & Gibberish detector 2 applied to $Dom-WDS$ \\
&$GIB-2-Dom-W2$  & Gibberish detector 2 applied to $Dom-W2$\\
&$GIB-2-Dom-W3$  & Gibberish detector 2 applied to $Dom-W3$ \\
\midrule
\multirow{5}{*}{Entropy}&$E-Dom$  & Entropy of  $Dom$ \\
&$E-Dom-WS$  & Entropy of $Dom-WS$ \\
&$E-Dom-D$  &  Entropy of $Dom-D$ \\
&$E-Dom-WDS$  & Entropy of $Dom-WDS$ \\
&$E-Dom-W2$  &  Entropy of $Dom-W2$ \\
&$E-Dom-W3$  &  Entropy of $Dom-W3$  \\
\bottomrule
\end{tabular}
%}
\end{table}
\endgroup

Using the insights from our analysis of DGA families in the dataset, several features were engineered, defined in \autoref{tab:parameters}.
The first set of parameters is computed when trying to identify valid n-grams and words.
For the former, we train our n-gram model with Alexa n-grams and lengths three, four and five. For the latter, the \texttt{wordninja}\footnote{https://github.com/keredson/wordninja} word splitter was used, which probabilistically analyses its input using NLP based on the unigram frequencies of the English Wikipedia. %...since domain names are usually concatenated words,
Hence, the domain is split into meaningful words, according to a minimum word-length $w$. Therefore, only terms which contain at least $w$ characters are considered as significant.
Then, we compute the percentage of the domain characters which are meaningful, by calculating the ratio $\gamma$ between characters belonging to words and the domain's total length.
Next, two more sets of features are computed according to statistical attributes as well as ratios using the previously calculated features.

\myparagraph{Gibberish Detection}
In addition, a Gibberish detection layer is used, which consists of two methods.
The first one is a 2-character Markov chain Gibberish detector\footnote{https://github.com/rrenaud/Gibberish-Detector}, which is trained with English text to determine how often characters appear next to each other.
Therefore, a text string is considered valid if it obtains a value above a certain threshold for each pair of characters.
The second is a Gibberish classifier.\footnote{https://github.com/ProgramFOX/GibberishClassifier-Python}
In this case, the method checks mainly three features of the text: whether (i) the amount of unique characters is within a typical range, (ii) the number of vowels is within a standard range and (iii), the word to char ratio is in a healthy range.
Finally, the entropy of a subset of the alphanumeric sequences is computed, to enrich the feature set.

\begin{figure}[th]
    \centering
    \includegraphics[width=\textwidth,trim=0in 1.65in 0in 0in, clip,keepaspectratio]{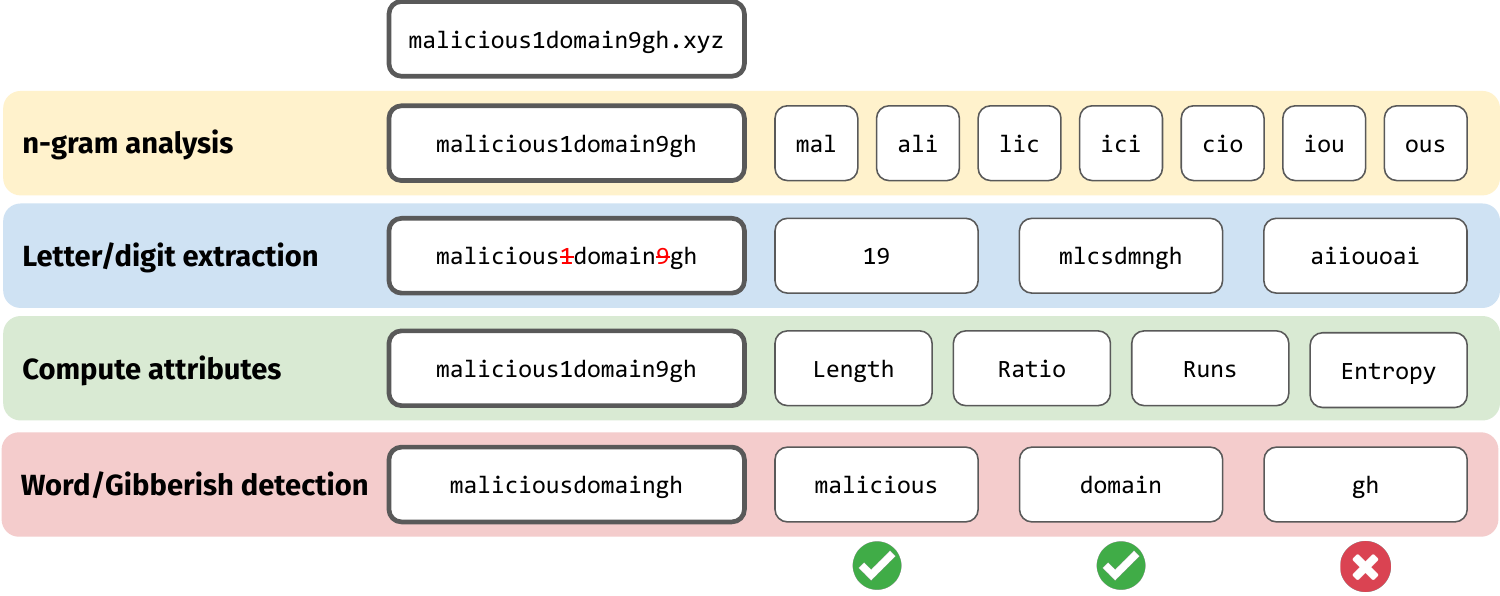}
    \caption{Exemplified overview of the feature extraction process.}
    \label{fig:featuregraph}
\end{figure}

An exemplified overview of the feature extraction process is illustrated in \autoref{fig:featuregraph}.

\myparagraph{A Comparison with the State-of-the-Art}
Despite the fact that n-grams and some of the ratio features used in this paper are well-known and have been previously used in the literature, the combination presented in this work is novel. Moreover, we propose the use of two different Gibberish detectors, the vowel distribution of the specific n-grams computed from the Alexa domains, the statistical features computed over the different length-based words extracted by \texttt{wordninja}, and the entropy used in a subset of this novel features.

%\section{Experimental Setup}
%\label{sec:experiments}
\section{Classification Experiments}\label{sec:classification}

We assess the power of our proposed features (see \autoref{sec:proposed-features}) to differentiate between malicious and benign domains (i.e., binary classification), as well as between several families of DGAs (i.e., multiclass classification).
Since both empirical and theoretical results have shown that a combination of models (in an ensemble) can increase classification performance~\cite{dietterich2000ensemble,valentini2002ensembles,barandela2003new,kuncheva2014combining} even in the case of imbalanced datasets~\cite{barandela2003new}, we opt for a Random Forest, which is a non-parametric ensemble classifier.
Random Forest has previously achieved outstanding performance results in DGA classification tasks~\cite{ALAEIYAN2020,Anderson2016,SELVI2019156}.

The hyperparameters of the Random Forest algorithm were tuned with grid search, to maximise classification performance in the task of distinguishing between benign and malicious domains over a subset of our dataset.
We found that best performance is achieved using an ensemble of 100 decision trees with unlimited depth and bootstrap aggregation (i.e., bagging), where each new tree is fitted from a bootstrap sample of the training data~\cite{breiman1996bagging}.

All our experiments were performed on a system equipped with an NVIDIA TITAN Xp PG611-c00 to speed-up the computations, utilising the \texttt{scikit-learn}\footnote{\url{https://scikit-learn.org}} library. The performance of the trained classifiers is evaluated using the standard classifications metrics of Precision, Recall, $F_1$ score and the area under the curve (AUC).%,

In all experiments,  the same feature set\footnote{We note that, on occasion, a few peculiarities might eliminate one feature due to reasons further detailed.} were used and employed standard 10-fold cross-validation to avoid overfitting and get a roughly unbiased estimate of the performance of the trained models. % for performance comparison purposes.
Although  the same feature sets for all experiments were used, we optimized weights of the features per DGA family, and thus targeting a binary classification (i.e., per DGA detection).
In a binary classification, this is a justified setting since feature weights for a particular DGA family are not expected to vary in time.
However, this is opposed to multi-class classification (requiring frequent feature/weight tuning), which we argue is not convenient for DGA detection since it deals with the more challenging classification problem.\footnote{I.e., intuitively, it is more difficult to find accurate separating hyperplanes among multiple classes than between two classes.}
Also, it should be noted that the multiclass classification is not the focus of this work, and those experiments carried out only for the sake of comparison with state-of-the-art approaches.

The set of experiments ran in this work aims to provide a solid proof of performance and accuracy for our approach. First, the detailed outcomes when applied to the HYDRAS dataset are provided. Next, we select two well-known state-of-the-art proposals using a method similar to ours, also based on Random Forest and implementing their own set of features. We compared the performance of such solutions to that of our method by applying them to the HYDRAS dataset.

\subsection{Binary Classification using the HYDRAS Dataset}
%We performed the binary classification using the Random Forest classifier. % IH: already mentioned above
%

In the current experiment, several binary Random Forest classifiers that correspond to DGA family detectors were cross-validated -- each detector is represented by a single such classifier.
To build an input sub-dataset of each DGA family detector, random sampling was employed without replacement on AGDs from the corresponding family (or benign samples) to fit a 1:1 ratio with the benign domains, resulting in a balanced sub-dataset.
To ensure the statistical significance of the results, each cross-validation execution was repeated 100 times (with different randomly selected sample subsets).
% robustness of the classifiers trained with our features.  \ih{robustness of classifiers is more observation that should be reported later}
%
In detail, for each DGA family in our dataset we ensured 1:1 ratio with benign domains, with the dataset size per each DGA detector of family $f$ equal to:
\begin{equation}
2 ~min( |H[f]|, |A|),
\end{equation}
where $A$ represents samples of Alexa dataset and $H[f]$ represents samples of a particular family $f$ in the \texttt{HYDRAS} dataset $H$.
Hence, in the case that the number of samples in $H[f]$ is greater than in $A$, we employ random sampling without replacement (across repeated experiments) on $H[f]$ to reduce its size to the size of~$A$.
In the opposite case, when the size of $A$ is greater than the size of $H[f]$, the same random sampling is employed to reduce the size of $A$, ensuring 1:1 ratio.
Note that for each of the 100 runs of cross-validation, different sample sets were randomly selected from the Alexa dataset as benign class representatives.\footnote{Nevertheless, for thorough evaluation, different ratios to reproduce the actual binary classification experiment will be later used (see \autoref{sec:discussion}).
}
Also, note that due to the particular format in which several families are available in their reverse-engineered form, as well as in the AGD repositories used to initialise our dataset, the $L-DOT$ feature could not be used homogeneously, and thus, was excluded from this experiment.

%\subsubsection{Summary}
The averaged outcomes of the binary classification can be seen in \autoref{tab:binary_stats}.
We can observe that most of the DGA families are classified almost without errors, obtaining precision and recall metrics of above 99.9\%.
Even in the case of families with small representation (e.g., \texttt{darkshell}, \texttt{omexo}, \texttt{qhost}, and \texttt{ud3}), the classifier can discern between benign and malicious domains in almost 100\% of cases, with only a few exceptions. Moreover, the standard deviation of the $F_1$ (i.e., $\sigma_{F_1}$) achieves values $\sim$1\% (in most cases only $\leq$0.01\%), which showcases the robustness of both the classifier and the proposed feature set.

The lowest accuracy was obtained by \texttt{bigviktor} (with a precision of 91.07\% and a recall of 76.44\%) followed by \texttt{suppobox}, \texttt{gozi, matsnu, khaos} and \texttt{symmi} with $F_1$ scores ranging between 95\% and 98\%.
This is due to the fact that most of these families use a composition of English dictionary words to create AGDs, so the extracted lexical features are not always able to properly differentiate them from our benign dataset or are adversarial.
In the case of such dictionary-based families, a further enhancement based on probabilistic-based methodologies\footnote{Note that these methodologies have demonstrated remarkable detection performance~\cite{patsakis2019exploiting}.} is an interesting but challenging future research direction. In the case of adversarially designed DGAs, the accuracy obtained is close to the one reported for the dictionary-based DGAs, which showcases the difficulty of capturing such families. A more detailed comparison and analysis of adversarially designed DGAs is later presented in \autoref{sec:adversarial}.
Overall, the outstanding detection rates showed in \autoref{tab:binary_stats} by using the same feature set across such a big dataset proves the robustness and adaptability of our approach.
Note that the more divergent families (and samples) are, the more difficult is to select a common set of features which can capture them accurately.

\begin{table}[t]
\vspace{-1.2cm}
\setlength{\tabcolsep}{3pt}

\def\arraystretch{1.1}
\centering
\caption{Performance measures for binary classification (in percentage).}
\label{tab:binary_stats}
\resizebox{\textwidth}{!}{
\begin{tabular}{>{\bfseries}lccccc>{\bfseries \columncolor{gray!25}}l>{\columncolor{gray!25}}c>{\columncolor{gray!25}}c>{\columncolor{gray!25}}c>{\columncolor{gray!25}}c>{\columncolor{gray!25}}c>{\bfseries}lccccc}
\hline
\textbf{Class}       & \textbf{Prec.}  & \textbf{Recall} & \textbf{$F_1$}   &\textbf{$\sigma_{F_1}$}    & \textbf{AUC}     & \textbf{Class}               &  \textbf{Prec.} & \textbf{Recall}                     & \textbf{$F_1$}    &\textbf{$\sigma_{F_1}$}    & \textbf{AUC}    & \textbf{Class}                   &  \textbf{Prec.} & \textbf{Recall} & \textbf{$F_1$}     &\textbf{$\sigma_{F_1}$}   & \textbf{AUC}    \\
\hline
bamital                                               & 100   & 100   & 100   & 0               & 100   & gspy                                       & 99.67 & 100   & 99.83 & 0.29            & 100   & qakbot                                       & 100   & 99.97 & 99.98 & 0.01            & 99.99 \\
banjori                                               & 99.99 & 99.74 & 99.87 & 0.01            & 99.87 & hesperbot & 100   & 99.85 & 99.92 & 0.01            & 99.93 & qhost                                        & 100   & 100   & 100   & 0               & 100   \\
bedep                                                 & 100   & 100   & 100   & 0               & 100   & infy                                       & 100   & 99.97 & 99.98 & 0.01            & 99.98 & qsnatch                                      & 99.77 & 99.86 & 99.81 & 0.02            & 99.81 \\
beebone                                               & 100   & 99.07 & 99.53 & 0.40             & 99.54 & khaos  & 99.47 & 96.47 & 97.95 & 0.10             & 97.98 & ramdo                                        & 100   & 100   & 100   & 0               & 100   \\
bigviktor                                             & 91.07 & 76.44 & 83.11 & 0.32            & 87.85 & kingminer & 100   & 97.62 & 98.8  & \textless{}0.01 & 98.81 & ramnit                                       & 100   & 99.95 & 99.97 & 0.02            & 99.98 \\
blackhole                                             & 100   & 99.86 & 99.93 & \textless{}0.01 & 99.93 & locky                                      & 100   & 99.79 & 99.9  & 0.04            & 99.9  & ranbyus                                      & 100   & 99.99 & 100   & 0               & 100   \\
\specialcell{bobax/\\~~/kraken\\~~/oderoor} & 100   & 99.62 & 99.81 & 0.01            & 99.81 & madmax    & 100   & 99.98 & 99.99 & \textless{}0.01 & 99.99 &redyms      & 100   & 99.63 & 99.82 & 0.32            & 99.82 \\
ccleaner            & 100   & 100   & 100   & 0               & 100   & makloader                                  & 100   & 100   & 100   & 0               & 100   & rovnix                                       & 100   & 100   & 100   & 0               & 100   \\
chinad                                                & 100   & 100   & 100   & 0               & 100   & matsnu                                     & 95.73 & 97.74 & 96.72 & 0.2             & 96.69 & shifu     & 100   & 99.63 & 99.82 & 0.01            & 99.82 \\
chir                                                  & 100 & 100   & 100 & 0            & 100   & mirai   & 100   & 99.96 & 99.98 & \textless{}0.01 & 99.98 & \specialcell{shiotob/\\~~/urlzone\\~~/bebloh}                       & 100   & 99.99 & 99.99 & \textless{}0.01 & 99.99 \\
conficker                                             & 99.99 & 99.64 & 99.81 & 0.02            & 99.82 & modpack                                    & 100   & 100   & 100   & 0               & 100   & simda      & 99.99 & 99.66 & 99.83 & 0.01            & 99.83 \\
corebot                                               & 100   & 99.98 & 99.99 & \textless{}0.01 & 99.99 & monerodownloader                           & 100   & 100   & 100   & 0               & 100   & sisron                                       & 100   & 100   & 100   & 0               & 100   \\
cryptolocker                                          & 100   & 99.98 & 99.99 & \textless{}0.01 & 99.99 & monerominer                                & 100   & 100   & 100   & 0               & 100   & sphinx                                       & 100   & 100   & 100   & 0               & 100   \\
cryptowall                                            & 100   & 99.87 & 99.93 & 0.02            & 99.94 & murofet                                    & 100   & 99.99 & 100   & \textless{}0.01 & 100   & suppobox    & 96.84 & 98.3  & 97.57 & 0.02            & 97.55 \\
darkshell                                             & 100   & 97.5  & 98.73 & 0               & 98.75 & murofetweekly                              & 100   & 100   & 100   & 0               & 100   & sutra                                        & 100   & 99.97 & 99.98 & \textless{}0.01 & 99.98 \\
deception         & 99.03 & 97.00    & 98.00    & 0.07            & 98.02 & mydoom                                     & 100   & 99.6  & 99.8  & 0.01            & 99.8  & symmi                                        & 100   & 93.85 & 96.83 & \textless{}0.01 & 96.92 \\
deception2        & 98.25 & 96.15 & 97.19 & 0.1             & 97.22 & necurs                                     & 100   & 99.89 & 99.95 & 0.02            & 99.95 & szribi                                       & 99.98 & 99.68 & 99.83 & 0.03            & 99.83 \\
diamondfox                                            & 100   & 97.13 & 98.55 & \textless{}0.01 & 98.57 & nymaim                                     & 100   & 99.6  & 99.8  & 0.03            & 99.8  & tempedreve  & 100   & 99.56 & 99.78 & 0.02            & 99.78 \\
dircrypt             & 100   & 99.94 & 99.97 & \textless{}0.01 & 99.97 & nymaim2   & 98.35 & 97.99 & 98.17 & 0.07            & 98.17 & tinba                                        & 100   & 99.92 & 99.96 & 0.02            & 99.96 \\
dmsniff                                               & 100   & 95.71 & 97.81 & \textless{}0.01 & 97.86 & omexo                                      & 100   & 100   & 100   & 0               & 100   & tinynuke                                     & 100   & 100   & 100   & 0               & 100   \\
dnschanger                                            & 100   & 99.93 & 99.96 & 0.02            & 99.97 & padcrypt                                   & 100   & 100   & 100   & 0               & 100   & tofsee                                       & 99.94 & 99.92 & 99.93 & 0.02            & 99.95 \\
dromedan            & 100   & 100   & 100   & 0               & 100   & pandabanker                                & 100   & 100   & 100   & 0               & 100   & torpig                                       & 100   & 99.79 & 99.89 & 0.01            & 99.9  \\
dyre                                                  & 100   & 100   & 100   & 0               & 100   & pitou                                      & 100   & 99.89 & 99.94 & 0.02            & 99.95 & tsifiri                                      & 100   & 100   & 100   & 0               & 100   \\
ebury                                                 & 100   & 100   & 100   & 0               & 100   & pizd     & 99.43 & 99.62 & 99.52 & 0.09            & 99.52 & ud2                                          & 100   & 100   & 100   & 0               & 100   \\
ekforward                                             & 100   & 100 & 100   & 0 & 100   & post                                       & 100   & 100   & 100   & 0               & 100   & ud3                                          & 100   & 100   & 100   & 0               & 100   \\
emotet                                                & 100   & 100   & 100   & 0               & 100   & proslikefan                                & 100   & 99.63 & 99.81 & 0.04            & 99.82 & ud4                                          & 100   & 96.19 & 98.06 & 0.43            & 98.1  \\
enviserv                                              & 100   & 100   & 100   & 0               & 100   & pushdo                                     & 99.94 & 98.99 & 99.46 & 0.02            & 99.46 & vawtrak    & 99.92 & 99.44 & 99.68 & 0.01            & 99.68 \\
feodo                                                 & 100   & 100   & 100   & 0               & 100   & pushdotid                                  & 100   & 99.62 & 99.81 & 0               & 99.81 & vidro    & 100   & 99.78 & 99.89 & 0.03            & 99.89 \\
fobber                                                & 100   & 99.85 & 99.92 & \textless{}0.01                & 99.93 & pykspa                                     & 100   & 99.7  & 99.85 & 0.01            & 99.85 & vidrotid                                     & 100   & 96.04 & 97.98 & \textless{}0.01 & 98.02 \\
fobber\_v1                                            & 100   & 100   & 100   & 0               & 100   & pykspa\_v1                                 & 100   & 99.28 & 99.64 & 0               & 99.64 & virut                                        & 99.97 & 99.99 & 99.98 & \textless{}0.01 & 99.98 \\
fobber\_v2                                            & 100   & 99.78 & 99.89 & 0.1             & 99.89 & pykspa\_v2\_fake                           & 100   & 99.77 & 99.89 & 0.01            & 99.89 & volatilecedar                                & 99.93 & 100   & 99.97 & 0.06            & 100   \\
gameover                                              & 100   & 100   & 100   & 0               & 100   & pykspa\_v2\_real                           & 100   & 99.77 & 99.88 & 0.01            & 99.88 & wd                                           & 100   & 100   & 100   & 0               & 100   \\
geodo                                                 & 100   & 100   & 100   & 0               & 100   & pykspa2                                    & 100   & 99.12 & 99.56 & 0.06            & 99.56 & xshellghost & 100   & 99.93 & 99.97 & 0.01            & 99.97 \\
gozi               & 95.28 & 95.93 & 95.6  & 0.11            & 95.59 & pykspa2s                                   & 100   & 97.47 & 98.72 & \textless{}0.01 & 98.74 & xxhex                                        & 100   & 99.96 & 99.98 & 0.02            & 99.98 \\
goznym                                                & 100   & 99.27 & 99.63 & 0.16            & 99.63 & qadars                                     & 100   & 99.98 & 99.99 & 0.01            & 99.99 & zloader     & 100   & 100   & 100   & 0               & 100  \\

\bottomrule
\end{tabular}
}
\end{table}

\begingroup
\medmuskip=0mu
\thickmuskip=0mu

\myparagraph{Feature Weighting}
For the sake of clarity, the average feature weight is depicted in the binary classification task in \ref{sec:feat_rel} and \autoref{tab:feature_w}.
The features exhibiting a high influence on the binary classification are $R-Dom-4G$, $R-Dom-5G$, $R-WS-LEN$, $R-WDS-LEN$, $GIB-1-Dom-D$, and $L-LEN$. Therefore, the n-grams, as well as the length of the domain and the valid words it contains, seem to be the most relevant features. % to distinguish between benign and malicious AGDs.
Besides that, the relevance of each feature varies according to the family. %A complete report on the per family weight of each feature can be found in~\cite{hydra_dataset}.
Feature weights provide some insights into how to try to further enhance the performance of our method, e.g., by employing various feature selection methods to reduce the number of features, which might be convenient in power-constrained devices such as IoT device or smartphones. %  computational power is scarce or the energy is a critical factor  ... and here it goes our next paper...
%Ha, ha, ha...(Julio) good plan!

\subsection{Binary Classification -- Comparison with State-of-the-Art Approaches}
In this experiment, we compare the quality of our features and classification approach
%and how they were able to capture the structure of sophisticated DGA families, which use dictionaries or adversarially-crafted AGDs. % IH: here is too early to claim us as winners
with two well-known approaches from the literature, namely the one proposed by Choudhary et al. \cite{choudhary2018algorithmically} and the method leveraged by Woodbridge et al. \cite{2}. Therefore, each feature set was implemented according to the corresponding specifications, and applied them to the HYDRAS dataset. Thereafter, the same setup was used than in our binary classification experiment and computed the $F_1$-score for each method. The outcome of this experiment is reported in \autoref{tab:multi_binary_comp}.

As it can be observed, all proposals succeed in capturing most of the families. The statistical features and n-gram-based features used in the different methods can recognise the patterns in the AGDs which belong to classical DGA families (i.e. the old ones which exhibit random-based generation patterns). In the case of more sophisticated families such as \texttt{rovnix, volatilecedar, beebone, banjori, locky, pushdo, proslikefan}, \texttt{symmi}, \texttt{goznym} and \texttt{pykspa}, all methods succeed to differentiate them from benign domains. Nevertheless, the method of Choudhary et al. reported notably worse accuracies for \texttt{pushdo, proslikefan, banjori, beebone, symmi, pykspa} and the variants of \texttt{pyskpa} than the method of Woodbridge et al. and our approach.

In the case of the rest of dictionary-based families, the adversarially-generated DGAs, and other novel DGA families, our method clearly outperforms the other approaches in most cases, with exception of \texttt{bigviktor}, in which the outcomes obtained by all methods are similar. For instance, in the case of \texttt{simda, vawtrak, gozi} and \texttt{qsnatch}, our method obtains a $F_1$ 3\% higher than the rest of methods.

In the case of more sophisticated dictionary-based families such as \texttt{matsnu, nymaim} and \texttt{suppobox}, our method outperforms the rest of approaches by approximately 10\% in the case of \texttt{matsnu}, and close to 18\% in the case of \texttt{nymaim} and \texttt{suppobox}. Finally, the highest difference was observed in the comparison with the adversarially generated DGAs.

In this regard, for the \texttt{deception} DGA, our approach outperforms Choudhary et al. by a 33\% and Woodbridge et al. by a 12\%. The \texttt{deception2} family exhibits similar outcomes, yet this time our approach outperforms Choudhary et al. by a 40\% and Woodbridge et al. by a 15\%. Finally, for the \texttt{khaos} DGA, the differences are close to 38\% when comparing to Choudhary et al. and approximately 12\% in the case of Woodbridge et al.

\begin{table}[!th]
\vspace{-1.8cm}
\centering
\def\arraystretch{1.0}
\caption{A comparison with state-of-the art (binary classification).} % using the $F_1$ score metric
\vspace{-0.3cm}

\rotatebox{90}{
\label{tab:multi_binary_comp}

\resizebox{\textheight}{!}{

\begin{tabular}{>{\bfseries}lcccccc>{\bfseries \columncolor{gray!25}}l>{\columncolor{gray!25}}c>{\columncolor{gray!25}}c>{\columncolor{gray!25}}c>{\columncolor{gray!25}}c>{\columncolor{gray!25}}c>{\columncolor{gray!25}}c>{\bfseries}lcccccc}
\hline

& \multicolumn{2}{c}{\textbf{\specialcell{Choudhary\\~et al.~\cite{choudhary2018algorithmically}}}}  & \multicolumn{2}{c}{\textbf{\specialcell{Woodbridge\\~~et al.~\cite{2}} }} &\multicolumn{2}{c}{\textbf{\specialcell{~~~Our\\Method}}}  & &

\multicolumn{2}{c}{\cellcolor{gray!25}\textbf{\specialcell{Choudhary\\~et al.~\cite{choudhary2018algorithmically}}}}  & \multicolumn{2}{c}{\cellcolor{gray!25}\textbf{\specialcell{Woodbridge\\~~et al.~\cite{2}}}} &\multicolumn{2}{c}{\cellcolor{gray!25}\textbf{\specialcell{~~~Our\\Method}}} & &

\multicolumn{2}{c}{\textbf{\specialcell{Choudhary\\~et al.~\cite{choudhary2018algorithmically}}}}  & \multicolumn{2}{c}{\textbf{\specialcell{Woodbridge\\~~et al.~\cite{2}}}} &\multicolumn{2}{c}{\textbf{\specialcell{~~~Our\\Method}}} \\
\textbf{Class} & \textbf{$F_1$}    &\textbf{$\sigma_{F_1}$} & \textbf{$F_1$}    &\textbf{$\sigma_{F_1}$} & \textbf{$F_1$}    &\textbf{$\sigma_{F_1}$} & \textbf{Class} & \textbf{$F_1$}    &\textbf{$\sigma_{F_1}$} & \textbf{$F_1$}    &\textbf{$\sigma_{F_1}$} & \textbf{$F_1$}    &\textbf{$\sigma_{F_1}$} & \textbf{Class} & \textbf{$F_1$}    &\textbf{$\sigma_{F_1}$} & \textbf{$F_1$}    &\textbf{$\sigma_{F_1}$} & \textbf{$F_1$}    &\textbf{$\sigma_{F_1}$}  \\
\midrule
bamital              & 99.99                & \textless{}0.01      & 99.99                & \textless{}0.01      & 100                  & 0                    & gspy             & 99.82                & 0.32                 & 100                  & 0                    & 99.83                & 0.29                 & qakbot                 & 97.93                & 0.17                 & 99.98                & 0.02                 & 99.98                & 0.01                 \\
banjori              & 96.53                & 0.04                 & 99.75                & 0.01                 & 99.87                & 0.01                 & hesperbot        & 97.18                & 0.09                 & 99.91                & 0.02                 & 99.92                & 0.01                 & qhost                  & 98.61                & 2.41                 & 97.19                & 1.18                 & 100                  & 0                    \\
bedep                & 99.15                & 0.03                 & 100                  & 0                    & 100                  & 0                    & infy             & 99.65                & 0.11                 & 99.98                & \textless{}0.01      & 99.98                & 0.01                 & qsnatch                & 95.15                & 0.14                 & 96.38                & 0.09                 & 99.81                & 0.02                 \\
beebone              & 97.05                & 4.52                 & 98.86                & 0.78                 & 99.53                & 0.4                  & khaos            & 59.68                & 0.56                 & 85.78                & 0.18                 & 97.95                & 0.1                  & ramdo                  & 99.84                & 0.02                 & 99.99                & \textless{}0.01      & 100                  & 0                    \\
bigviktor            & 80.82                & 0.93                 & 85.47                & 0.79                 & 83.11                & 0.32                 & kingminer        & 99.40                & 0.2                  & 99.93                & 0.11                 & 98.8                 & \textless{}0.01      & ramnit                 & 97.25                & 0.18                 & 99.95                & 0.02                 & 99.97                & 0.02                 \\
blackhole            & 99.82                & 0.14                 & 99.93                & \textless{}0.01      & 99.93                & \textless{}0.01      & locky            & 95.18                & 0.14                 & 99.84                & 0.01                 & 99.99                & 0.04                 & ranbyus                & 99.54                & 0.02                 & 99.99                & \textless{}0.01      & 100                  & 0                    \\
\specialcell{bobax/\\~~/kraken\\~~/oderoor} & 95.18                & 0.09                 & 99.74                & 0.03                 & 99.81                & 0.01                 & madmax           & 99.00                & 0.05                 & 99.99                & \textless{}0.01      & 99.99                & \textless{}0.01      & redyms                 & 98.37                & 0.54                 & 99.46                & 0.94                 & 99.82                & 0.32                 \\
ccleaner             & 99.87                & 0.01                 & 100                  & 0                    & 100                  & 0                    & makloader        & 99.94                & 0.11                 & 99.94                & 0.11                 & 100                  & 0                    & rovnix                 & 99.97                & 0.01                 & 99.99                & \textless{}0.01      & 100                  & 0                    \\
chinad               & 99.95                & \textless{}0.01      & 99.99                & \textless{}0.01      & 100                  & 0                    & matsnu           & 85.77                & 0.37                 & 85.74                & 0.26                 & 96.72                & 0.2                  & shifu                  & 97.15                & 0.20                 & 99.52                & 0.05                 & 99.82                & 0.01                 \\
chir                 & 100                  & 0                    & 100                  & 0                    & 100                  & 0                    & mirai            & 99.09                & 0.09                 & 100                  & 0                    & 99.98                & \textless{}0.01      & \specialcell{shiotob/\\~~/urlzone\\~~/bebloh} & 99.10                & 0.02                 & 99.99                & 0.01                 & 99.99                & \textless{}0.01      \\
conficker            & 93.02                & 0.16                 & 99.02                & 0.08                 & 99.81                & 0.02                 & modpack          & 96.49                & 1.47                 & 99.92                & 0.13                 & 100                  & 0                    & simda                  & 95.22                & 0.19                 & 95.04                & 0.28                 & 99.83                & 0.01                 \\
corebot              & 99.78                & 0.05                 & 99.99                & 0.01                 & 99.99                & \textless{}0.01      & \specialcell{monero-\\~~downloader} & 99.91                & 0.02                 & 100                  & 0                    & 100                  & 0                    & sisron                 & 99.86                & 0.04                 & 100                  & 0                    & 100                  & 0                    \\
cryptolocker         & 98.87                & 0.02                 & 99.99                & 0.01                 & 99.99                & \textless{}0.01      & monerominer      & 99.95                & 0.01                 & 100                  & 0                    & 100                  & 0                    & sphinx                 & 99.73                & 0.02                 & 99.99                & 0.01                 & 100                  & 0                    \\
cryptowall           & 97.13                & 0.17                 & 99.96                & 0.01                 & 99.93                & 0.02                 & murofet          & 99.3                 & 0.06                 & 100                  & 0                    & 100                  & \textless{}0.01      & suppobox               & 79.64                & 0.42                 & 80.32                & 0.38                 & 97.57                & 0.02                 \\
darkshell            & 99.18                & 0.71                 & 99.59                & 0.71                 & 98.73                & 0                    & murofetweekly    & 99.99                & \textless{}0.01      & 100                  & 0                    & 100                  & 0                    & sutra                  & 99.73                & 0.07                 & 99.98                & \textless{}0.01      & 99.98                & \textless{}0.01      \\
deception            & 64.96                & 0.52                 & 86.31                & 0.19                 & 98.00                & 0.07                 & mydoom           & 98.61                & 0.15                 & 99.80                & 0.03                 & 99.80                & 0.01                 & symmi                  & 90.46                & 1.05                 & 97.42                & 0.43                 & 96.83                & \textless{}0.01      \\
deception2           & 57.80                & 0.27                 & 82.14                & 0.49                 & 97.19                & 0.10                 & necurs           & 96.84                & 0.09                 & 99.94                & 0.01                 & 99.95                & 0.02                 & szribi                 & 97.64                & 0.12                 & 99.58                & 0.04                 & 99.83                & 0.03                 \\
diamondfox           & 98.60                & 0.39                 & 99.39                & 0.21                 & 98.55                & \textless{}0.01      & nymaim           & 93.17                & 0.16                 & 98.99                & 0.12                 & 99.80                & 0.03                 & tempedreve             & 94.68                & 0.06                 & 99.54                & 0.05                 & 99.78                & 0.02                 \\
dircrypt             & 97.65                & 0.02                 & 99.97                & \textless{}0.01      & 99.97                & \textless{}0.01      & nymaim2          & 77.62                & 0.45                 & 80.01                & 0.52                 & 98.17                & 0.07                 & tinba                  & 99.23                & 0.05                 & 99.97                & 0.01                 & 99.96                & 0.02                 \\
dmsniff              & 96.24                & 1.15                 & 100                  & 0                    & 97.81                & \textless{}0.01      & omexo            & 100                  & 0                    & 99.60                & 0.70                 & 100                  & 0                    & tinynuke               & 99.99                & 0.01                 & 100                  & 0                    & 100                  & 0                    \\
dnschanger           & 98.64                & 0.03                 & 99.97                & 0.01                 & 99.96                & 0.02                 & padcrypt         & 99.82                & 0.03                 & 99.99                & \textless{}0.01      & 100                  & 0                    & tofsee                 & 99.92                & 0.03                 & 99.88                & 0.12                 & 99.93                & 0.02                 \\
dromedan             & 98.53                & 0.04                 & 100                  & 0                    & 100                  & 0                    & pandabanker      & 99.95                & 0.01                 & 100                  & 0                    & 100                  & 0                    & torpig                 & 95.80                & 0.14                 & 99.08                & 0.09                 & 99.89                & 0.01                 \\
dyre                 & 99.99                & \textless{}0.01      & 100                  & 0                    & 100                  & 0                    & pitou            & 99.22                & 0.01                 & 99.59                & 0.14                 & 99.94                & 0.02                 & tsifiri                & 100                  & 0                    & 100                  & 0                    & 100                  & 0                    \\
ebury                & 99.83                & 0.09                 & 100                  & 0                    & 100                  & 0                    & pizd             & 88.29                & 0.10                 & 88.27                & 0.26                 & 99.52                & 0.09                 & ud2                    & 99.97                & 0.06                 & 100                  & 0                    & 100                  & 0                    \\
ekforward            & 99.28                & 0.09                 & 99.96                & 0.01                 & 100                  & 0                    & post             & 99.99                & 0.01                 & 100                  & 0                    & 100                  & 0                    & ud3                    & 100                  & 0                    & 98.37                & 1.41                 & 100                  & 0                    \\
emotet               & 99.77                & 0.04                 & 99.99                & 0                    & 100                  & 0                    & proslikefan      & 94.13                & 0.14                 & 99.48                & 0.06                 & 99.81                & 0.04                 & ud4                    & 96.40                & 0.35                 & 100                  & 0                    & 98.06                & 0.43                 \\
enviserv             & 99.23                & 0.13                 & 99.93                & 0.12                 & 100                  & 0                    & pushdo           & 92.00                & 0.14                 & 97.52                & 0.06                 & 99.46                & 0.02                 & vawtrak                & 90.53                & 0.24                 & 95.56                & 0.18                 & 99.68                & 0.01                 \\
feodo                & 99.46                & 0.12                 & 100                  & 0                    & 100                  & 0                    & pushdotid        & 98.15                & 0.09                 & 99.80                & 0.04                 & 99.81                & 0                    & vidro                  & 95.77                & 0.19                 & 99.82                & 0.03                 & 99.89                & 0.03                 \\
fobber               & 98.64                & 0.17                 & 99.96                & \textless{}0.01      & 99.92                & \textless{}0.01      & pykspa           & 94.12                & 0.24                 & 99.51                & 0.03                 & 99.85                & 0.01                 & vidrotid               & 93.03                & 0.97                 & 97.06                & \textless{}0.01      & 97.98                & \textless{}0.01      \\
fobber\_v1           & 99.92                & 0.08                 & 100                  & 0                    & 100                  & 0                    & pykspa\_v1       & 93.80                & 0.51                 & 99.55                & 0.04                 & 99.64                & 0                    & virut                  & 97.02                & 0.16                 & 97.68                & 0.03                 & 99.98                & \textless{}0.01      \\
fobber\_v2           & 98.20                & 0.54                 & 100                  & 0                    & 99.89                & 0.10                 & pykspa\_v2\_fake & 96.41                & 0.14                 & 99.63                & 0.02                 & 99.89                & 0.01                 & volatilecedar          & 99.37                & 0.21                 & 99.83                & 0.15                 & 99.97                & 0.06                 \\
gameover             & 99.97                & 0.02                 & 100                  & 0                    & 100                  & 0                    & pykspa\_v2\_real & 95.60                & 0.19                 & 99.46                & 0.03                 & 99.88                & 0.01                 & wd                     & 99.99                & \textless{}0.01      & 100                  & 0                    & 100                  & 0                    \\
geodo                & 99.79                & 0.09                 & 99.96                & 0.03                 & 100                  & 0                    & pykspa2          & 92.63                & 0.87                 & 99.18                & 0.03                 & 99.56                & 0.06                 & xshellghost            & 98.63                & 0.08                 & 99.96                & 0.01                 & 99.97                & 0.01                 \\
gozi                 & 88.65                & 0.41                 & 92.45                & 0.27                 & 95.60                & 0.11                 & pykspa2s         & 92.21                & 0.29                 & 99.07                & 0.29                 & 98.72                & \textless{}0.01      & xxhex                  & 99.60                & 0.03                 & 100                  & 0                    & 99.98                & 0.02                 \\
goznym               & 92.89                & 0.77                 & 99.68                & 0.08                 & 99.63                & 0.16                 & qadars           & 99.70                & 0.08                 & 99.99                & \textless{}0.01      & 99.99                & 0.01                 & zloader                & 99.93                & 0.01                 & 100                  & 0                    & 100                  & 0                           \\
\hline
\end{tabular}
}
}

\end{table}

%\subsubsection{Summary}
The average $F_1$ measure per family, as well as the standard deviation of the total average, are depicted in \autoref{tbl:compbinavg}. As it can be observed, since a significant amount of DGA families are detected with high performance by all approaches, the average $F_1$ outcomes are high in all methods. Nevertheless, the robustness of our methodology is highlighted by the $\sigma$ value, since it translates into a very high detection rate across all families, outperforming the rest of approaches.

\begin{table}[t]
    \centering
     \caption{Average outcomes per DGA class in the binary classification comparison.}
     \scriptsize
    \begin{tabular}{lccc}
    \toprule
         &\textbf{Choudhary et al.~\cite{choudhary2018algorithmically} }& \textbf{Woodbridge et al.~\cite{2}}	& \textbf{Our Method}	\\
         \midrule
        \textbf{Average $F_1$}    & 96.019   & 98.323   &  99.454 \\
         \textbf{$\sigma$} & 7.424  & 4.289 & 1.810 \\
         \bottomrule
    \end{tabular}

    \label{tbl:compbinavg}
\end{table}

The aforementioned comparison supports our idea to use a novel and upgradeable dataset such as HYDRAS, since most of the families present in the datasets used in the state-of-the-art approaches belong either to the random-based DGA category or to the thoroughly analysed set of dictionary-based families with specific patterns. In both cases, such families can be captured with high classification performance by using well-known feature sets. The latter implies that researchers should evaluate their methods with novel more complex families since approaches that test their accuracy with old datasets are no longer proving their validity versus the current DGA landscape.

\subsection{Binary Classification using Other Datasets}
To compare the quality of our approach when applied to other datasets, two datasets from the recent literature are selected. First, we selected the dataset presented in~\cite{ANAND20201129}, which contains 50,600 samples that are split to benign and malicious AGDs in 50:50 ratio (i.e., 25,300 benign samples and 25,300 malicious ones). The authors used several machine learning methods and reported accuracy between 94.9\% and 97.0\% for the C5.0 algorithm -- the one that achieved the highest accuracy. In our case, using the same dataset, we achieved the accuracy of 98.9\%, which is between 1.9\% and 4\% higher than any of the proposed methods in~\cite{ANAND20201129}.
%'Precision': 0.992072929052715, 'Recall': 0.9881563363600474, 'F1': 0.9901107594936709 AUC: 0.9889269808313398 Accuracy: 0.988825067608341

The second comparison was done using the dataset created in~\cite{SELVI2019156}. In this case, the authors tested their approach with a dataset consisting of 64,000 samples -- similar to the previous case, the samples were split in 50:50 ratio for the benign and malicious classes. The authors of~\cite{SELVI2019156} used a Random Forest classifier and achieved an accuracy of 98.9\% by using their 2-gram setup with 34 features.
In our case, the accuracy achieved was 96.7\%, that is only 2\% below the outcome achieved by the authors.
Nevertheless, the main drawback of their approach is the computational time required to compute such n-grams, which grows exponentially. In fact, the authors of~\cite{SELVI2019156} needed 1.21h for execution of a complete n-fold experiment with three repetitions.
In our case, the same experiment took approximately one minute. The latter emphasises that the trade-off between accuracy and computational time is also an important aspect (see \autoref{sec:performance} for the details).
% the trade-off between accuracy and computational time is clearly outperformed by our approach.
%'Precision': 0.9849757673667205, 'Recall': 0.9563321517081281, 'F1': 0.9704426440861413 AUC: 0.9687321135899131 Accuracy: 0.9671455362513711

\subsection{Multiclass Classification using Other Datasets}
In this experiment, we
% further explore the potential of our proposed set of features to go beyond a binary classification and
aim at predicting the DGA families, given a set of AGDs, in a multiclass classification setting.
We argue that this experiment has low practical utility in contrast to binary classification, and is provided only due to a fair comparison with related work.

A fair comparison of the multiclass classification's performance cannot be made if different approaches use different datasets.
Although the best option to compare a performance of related work vs. our approach would be to use the \texttt{HYDRAS} dataset, several problems arise: 1) many implementations of feature extractors are unavailable, 2) classifiers need to be fine-tuned and details of parameters are often not presented in papers.
Intuitively, due to the a large number of DGA families contained in the HYDRAS dataset, the multiclass classification using HYDRAS dataset yields worse outcomes than in small datasets.
This is the common problem of multiclass classification, especially when the classes are not well separated \cite{silva2017improving}.
%
%There exist several research direction aiming to reduce the impact of high number of classes in this context by, e.g., using dimensionality reduction~\cite{song2008unified}, or using hierarchical approaches~\cite{silva2017improving}, yet these direction are left to future work\ih{this sentence seems wrong to me -- dimensionality reduction does not help since it concerns only reduction of features and thus course of dimensionality -- I suggest to drop it.}.

\begin{table}[t]
	\centering
	\rowcolors{2}{gray!25}{white}
	\def\arraystretch{1.1}

	\caption{Multiclass classification outcomes in percentages, using different performance metrics. The averages are weighted according to the number of samples in each family.}

	\label{tab:multi_stats}
	\resizebox{\textwidth}{!}{%
		\begin{tabular}{>{\bfseries}l|ccc|ccc|ccc|ccc}
			\hline
			& \multicolumn{3}{|c|}{\textbf{Phoenix~\cite{stefanotracking}}}  & \multicolumn{3}{|c|}{\textbf{DeepDGA~\cite{Anderson2016}}} & \multicolumn{3}{|c|}{\textbf{Alaeiyan et al.~\cite{ALAEIYAN2020}}} &\multicolumn{3}{|c}{\textbf{Our Method}}  \\

			\textbf{Class}       & \multicolumn{1}{|l}{\textbf{Precision}} & \multicolumn{1}{l}{\textbf{Recall}} & \multicolumn{1}{l|}{\textbf{$F_1$}} & \multicolumn{1}{|l}{\textbf{Precision}} & \multicolumn{1}{l}{\textbf{Recall}} & \multicolumn{1}{l|}{\textbf{$F_1$}} & \multicolumn{1}{|l}{\textbf{Precision}}                       & \multicolumn{1}{l}{\textbf{Recall}}                         & \multicolumn{1}{l|}{\textbf{$F_1$}} & \multicolumn{1}{|l}{\textbf{Precision}}                 & \multicolumn{1}{l}{\textbf{Recall}}                    & \multicolumn{1}{l}{\textbf{$F_1$}} \\
			\hline
			banjori     & 100                           & 100                        & 100                 & 100                           & 100                        & 100                 & 100                             & 100                            & 100                    & 99.80                     & 100                       & 99.90                  \\
			chinad      & 74.2                          & 84.1                       & 78.84                  & 98.8                          & 96.3                       & 97.53                  & 92.3                            & 19.5                           & 32.20                  & 87.99                     & 97.27                     & 92.39                  \\
			corebot     & 88.4                          & 97.4                       & 92.68                  & 100                           & 100                        & 100                    & 100                             & 97.4                           & 98.68                  & 100                       & 72.50                     & 84.06                  \\
			dircrypt    & 0                             & 0                          & 0                  & 0                             & 0                          & 0                      & 0                               & 0                              & 0                      & 0                         & 0                         & 0                      \\
			downloader  & 100                           & 100                        & 100                    & 100                           & 100                        & 100                    & 100                             & 100                            & 100                    & 100                       & 100                       & 100                    \\
			dnschanger  & 0                             & 0                          & 0                      & 0                             & 0                          & 0                      & 0                               & 0                              & 0                      & 0                         & 0                         & 0                      \\
			fobber      & 3.2                           & 2                          & 2.46                   & 3.1                           & 7.6                        & 4.40                   & 0                               & 0                              & 0                      & 37.62                     & 12.67                     & 18.95                  \\
			gozi        & 7.1                           & 50.0                         & 12.43                  & 3.0                             & 3.0                          & 3.00                   & 0                               & 0                              & 0                      & 0                         & 0                         & 0                      \\
			javascript  & 2.0                             & 3.4                        & 2.52                   & 0                             & 0                          & 0                      & 0                               & 0                              & 0                      & 0                         & 0                         & 0                      \\
			locky       & 0                             & 0                          & 0                      & 0                             & 0                          & 0                      & 0                               & 0                              & 0                      & 0                         & 0                         & 0                      \\
			murofet     & 99.9                          & 98.0                         & 98.94                  & 100                           & 98.0                         & 98.99                  & 99.7                            & 99.4                           & 99.55                  & 99.91                       & 99.79                       & 99.85                    \\
			necurs      & 22.2                          & 15.8                       & 18.46                  & 28.0                            & 32.7                       & 30.17                  & 30.2                            & 5.1                            & 8.73                   & 32.24                     & 8.15                      & 13.02                  \\
			newgoz      & 97.2                          & 94.1                       & 95.62                  & 100                           & 99.6                       & 99.80                  & 98                              & 89.9                           & 93.78                  & 99.40                     & 100                       & 99.70                  \\
			kraken      & 88.1                          & 52.3                       & 65.64                  & 90.7                          & 60.3                       & 72.44                  & 98.5                            & 90                             & 94.06                  & 99.30                     & 99.93                     & 99.61                  \\
			padcrypt    & 79.3                          & 100                        & 88.46                  & 88.5                          & 100                        & 93.90                  & 100                             & 65.2                           & 78.93                  & 87.50                     & 29.17                     & 43.75                  \\
			proslikefan & 3.0                             & 19.4                       & 5.20                  & 4.0                             & 23.5                       & 6.84                   & 0                               & 0                              & 0.00                   & 25.00                     & 2.00                      & 3.70                   \\
			pykspa      & 85.2                          & 61.4                       & 71.37                  & 89.1                          & 80.3                       & 84.47                  & 84.7                            & 99.4                           & 91.46                  & 83.50                     & 86.86                     & 85.15                  \\
			qadars      & 60.4                          & 81.1                       & 69.24                  & 85.2                          & 84.7                       & 84.95                  & 96.9                            & 16.3                           & 27.91                  & 69.70                     & 34.50                     & 46.15                  \\
			qakbot      & 53.5                          & 55.0                         & 54.24                  & 57.9                          & 56.6                       & 57.24                  & 54.5                            & 82.1                           & 65.51                  & 65.69                     & 76.70                     & 70.77                  \\
			ramnit      & 1.2                           & 3.3                        & 1.76                   & 0                             & 0                          & 0                      & 0                               & 0                              & 0                      & 0                         & 0                         & 0                      \\
			ranbyus     & 2.1                           & 6.5                        & 3.17                   & 0                             & 0                          & 0                      & 0                               & 0                              & 0                      & 0                         & 0                         & 0                      \\
			shiotob     & 96.7                          & 81.6                       & 88.51                  & 98.3                          & 89.8                       & 93.86                  & 84.7                            & 91.4                           & 87.92                  & 92.26                     & 90.00                     & 91.12                  \\
			simda       & 63.0                            & 99.0                        & 77.00                     & 98.3                          & 89.8                       & 93.86                  & 89.6                            & 100                            & 94.51                  & 91.33                     & 99.00                     & 95.01                  \\
			sisron      & 100                           & 100                        & 100                    & 100                           & 100                        & 100                    & 100                             & 100                            & 100                    & 100                       & 100                       & 100                    \\
			suppobox    & 32.4                          & 79.3                       & 46.00                  & 67.4                          & 74.6                       & 70.82                  & 97.3                            & 68.5                           & 80.40                  & 90.61                     & 98.43                     & 94.36                  \\
			symmi       & 98.3                          & 96.6                       & 97.44                  & 98.3                          & 100                        & 99.14                  & 98.3                            & 100                            & 99.14                  & 92.31                     & 56.25                     & 69.90                  \\
			tempedreve  & 27.6                          & 67.1                       & 39.11                  & 43.8                          & 96.3                       & 60.21                  & 57.2                            & 75.1                           & 64.94                  & 59.98                     & 71.94                     & 65.42                  \\
			tinba       & 25.3                          & 64.6                       & 36.36                  & 49.9                          & 98.2                       & 66.17                  & 100                             & 99.7                           & 99.85                  & 99.52                     & 99.94                     & 99.73                  \\
			vawtrak     & 30.9                          & 9.7                        & 14.77                  & 68.3                          & 87.5                       & 76.72                  & 100                             & 8.3                            & 15.33                  & 69.47                     & 66.00                     & 67.69                  \\
			\hline
			\textbf{Total}       & \textbf{93.25}                   & \textbf{90.46}                & \textbf{91.40}                  & \textbf{94.64}                   & \textbf{92.49}                 & \textbf{93.28}                  & \textbf{94.49} & \textbf{95.20} & \textbf{94.41}     & \textbf{95.39} & \textbf{95.49} & \textbf{95.25}       \\
			\hline
		\end{tabular}
	}

\end{table}

Therefore, this experiment is based on another dataset introduced in~\cite{badergithub}, which was already evaluated by several papers.
For example, this dataset was evaluated by Alaeiyan et al.~\cite{ALAEIYAN2020}.
We further extend the comparison with the other two approaches reported in the literature, namely DeepDGA~\cite{Anderson2016} and Phoenix~\cite{stefanotracking}.
We performed a multiclass classification using our Random Forest classifier and our proposed feature set, this time slightly changing its configuration from the one used for the binary classification (i.e., an ensemble of 200 trees with unlimited depth).
The repository of this dataset no longer contained samples of the \texttt{RunForestrun} family, so it was not included in the comparison.
Moreover, the \texttt{Tinba} family had repeated SLD entries, which would lead to a biased and possibly unrepresentative classification (i.e., the same samples could easily end up both in the training and testing sets, thus reporting a 100\% detection in most of the validations partly due to this fact).
Therefore, a subset of the \texttt{Tinba} samples that are presented in our dataset was used. It is worth to note that this issue was ignored or not reported by the rest of approaches using this dataset.

%\subsubsection{Summary}
The outcomes of the multiclass classification are depicted in \autoref{tab:multi_stats}.
Even though in some cases Phoenix and DeepDGA showed better performance (e.g., for \texttt{Padcrypt}, \texttt{Qadars}, and \texttt{Symmi}), our method outperformed the rest, both in accuracy and recall, followed by the method proposed in~\cite{ALAEIYAN2020}.
Furthermore, we observed that there were relatively small differences for most families, and for the most part all methods reported high accuracy for similar families.
In some cases, the reported performance metrics were equal to zero, which is related to the lack of samples. The latter means that the classifier could not be properly trained.

%With regard to the feature relevance,
We can observe in \autoref{tab:feature_w} of Appendix that feature relevance in the multiclass setting differs substantially from the reported in the binary classification.
%This indicates the sensitivity of the subset of families conforming the dataset.
In other words, the relevance of the features will be different according to the particularities of the collected samples, highlighting the importance of using the same dataset to avoid biased comparisons across different approaches. In the particular case of the dataset selected to perform the multiclass classification, the features $L-LEN$, $E-Dom$, $L-DIG$, $R-WD-LEN$, and $R-WDS-LEN$ were the most relevant.
The latter means that the length (i.e., specific families create only AGDs of fixed length), the number of digits, the meaningful words in the SLD, and the entropy of the SLD enabled to predict which specific family created a given AGD, with a high precision.
\endgroup

\section{Classification of Adversarially Designed AGDs}
\label{sec:adversarial}
To further assess the quality of our selected features, we opted to use it against three especially ``hard to detect'' DGAs. These DGAs, \texttt{deception}, \texttt{deception2}~\cite{Spooren2019}, and \texttt{khaos}~\cite{8936543} are specially crafted, using machine learning methods, to evade detection (see \autoref{sec:background}). While our features are generic and not targeted towards identifying any particular set of these families, we also manage to detect adversarially designed AGDs with significantly better performance than in previous works (see \autoref{tbl:adversarial}).
In detail, the precision achieved by our approach is by 15\% to 30\% better.
%check this!
Similarly, the recall and F1 score are by more than 10\% better in almost all cases.
%and this, and sell it better!
It may also be observed that in some cases, the detection rates slightly vary if the ratio of malicious to benign samples is increased. This fact will be discussed more thoroughly in \autoref{sec:discussion}. Nonetheless, the F1 score is at least 92.48\%, which indicates that our method is very effective even when confronted against specially crafted DGAs -- a challenge that is very close to represent the most challenging scenario.

\subsection{Invalid Domains}
It should be noted that, during our study,  many invalid domains generated by these DGAs were identified. To the best of our understanding, the researchers simply left the neural networks to generate AGDs that bypassed the filters, without double-checking their validity in real scenarios. As a result, the neural networks identified that the use of the hyphen character managed to bypass some filters, priming them to overuse it. Therefore, one can observe thousands of domains which either start or finish with a hyphen, which are unfortunately invalid according to RFC 1123~\cite{braden1989rfc1123}. Similarly, many of them do not conform to RFC 5891 for Internationalized Domain Names (IDNs)~\cite{klensin2010internationalized}, by having hyphens as third and fourth characters, but not starting with an ``xn'', so they are rejected by ICANN\footnote{\url{https://www.icann.org/en/system/files/files/idn-guidelines-10may18-en.pdf}}. The issue is particularly relevant in the DGAs generated by Spooren et al.~\cite{Spooren2019} spanning across 1.64\% of the samples. In the \texttt{khaos} family, the dataset contains 19 IDNs.
Note that in all following experiments,  only DGAs which do not produce IDNs are considered.

\subsection{Detection of Unknown Families}
We performed another experiment to test the capability of our approach to detect previously unknown DGA families. In this regard, a leave-one-out experiment were leveraged with the same configuration as in the binary classification experiments (i.e., 10-fold cross validation) but in contract to it the target family was completely hidden to the training phase. In other words, we tried to predict whether a set of AGDs is benign or malicious without previous knowledge of the DGA family generating them.
%Considering a binary classification between the \textit{benign} and the \textit{malicious} classes.
The outcome of this experiment is reported in \autoref{tbl:leaveoneout}. As it can be observed in the table, our approach is able to correctly classify most of the samples, achieving a slightly lower $F_1$-score than the one reported in our binary classification (see \autoref{tbl:adversarial}), due to a general decrease of the recall values. Note that the most affected family is \texttt{deception2}, yet our approach still outperforms the original works in which these families were proposed, thus showcasing the robustness of our features once again.

\begin{table}[t]
	\centering
	\caption{A detection of unknown DGA families represented by adversarially designed DGAs (leave-one-out experiment).}
	\footnotesize
	\begin{tabular}{lcccc}
		\toprule
		\textbf{DGA}& \textbf{Precision}	& \textbf{Recall}	&\textbf{F1}	\\
		\midrule
		khaos    & 100   &85.40   &  92.12 \\
		deception & 99.99  & 84.71 & 91.72 \\
		deception2  & 99.99  &  73.08& 84.44 \\
		\bottomrule
	\end{tabular}

	\label{tbl:leaveoneout}
\end{table}

\begin{table}[h]
	\centering
	\caption{Binary classification against adversarially designed DGAs. First row of each family denotes the reported results in the original work.}
	\footnotesize
	\begin{tabular}{lcccc}
		\toprule
		\textbf{DGA}	&\textbf{Method} & \textbf{Precision}	& \textbf{Recall}	&\textbf{F1}	\\
		\midrule
		\multirow{4}{*}{khaos}& Yun et al. \cite{8936543}&68.00&98.00&80.30\\
		& Our approach - ratio 1:1&	99.47&	96.47&	97.95\\
		&Our approach - ratio 1:10&	96.08&	90.73& 93.32	\\
		&Our approach - ratio 1:100&	96.55&	89.63	&92.96 \\
		\rowcolor{gray!25}& Spooren et al. \cite{Spooren2019}& 84.40 & 87.10 & 85.72 \\
		\rowcolor{gray!25}&	Our approach - ratio 1:1&	99.03&97.00		& 98.00\\
		\rowcolor{gray!25}&Our approach - ratio 1:10&	96.21&	93.86&	95.02\\
		\rowcolor{gray!25}\multirow{-4}{*}{deception}&Our approach - ratio 1:100&	96.12&	93.29& 94.68\\
		\multirow{4}{*}{deception2}& Spooren et al. \cite{Spooren2019} & 77.50  & 81.50  & 79.45\\
		&	Our approach - ratio 1:1&	98.25&	96.15&	97.19\\
		&Our approach - ratio 1:10&	94.44&	91.29	&92.84\\
		&Our approach - ratio 1:100&	94.56&	90.50	&92.48\\
		\bottomrule
	\end{tabular}

	\label{tbl:adversarial}
\end{table}

\section{Overhead of Our Approach}
\label{sec:performance}
\begin{figure}[t]
    \centering
    \includegraphics[width=.7\textwidth]{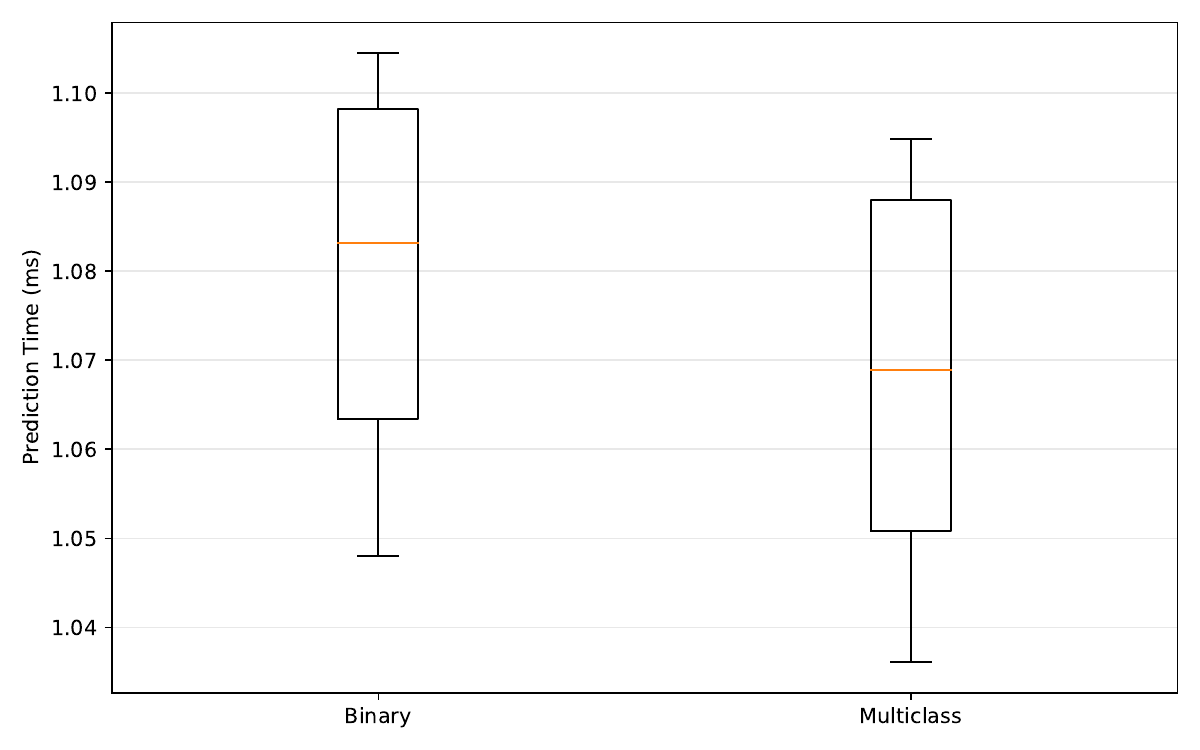}
    \caption{Prediction time per SLD, in milliseconds.}
    \label{fig:box_plot}
\end{figure}

We measured several statistics during our experiments with the intention to demonstrate the practical aspects of our approach.
We measured the duration of the features computation and the prediction time in both the binary and multiclass setting with our dataset.
The average time required to compute all the features for an SLD is 1.48 ms,
%will be good to offer here a +- standard deviation
while the prediction times from both classification experiments are depicted in \autoref{fig:box_plot}.
Note that the figure does not include the training time, which scales linearly with the size of the dataset. %\ih{maybe check training complexity of random forest and report it here}.
However, training is an action that is very rare (e.g., repeated after weeks or months) and it can be performed offline.
Hence, it does not incur performance degradation to the operation of AGD detector.
Finally, both the features computation time and the prediction times are measured without any parallelisation to enable a fair comparison with related work.

\subsection{Binary Classification}
In terms of performance comparison for the binary classification, the work proposed in \cite{ANAND20201129} did not report any computational cost nor performance metrics. In the case of \cite{SELVI2019156}, the authors reported a total experiment time of 1.21h in their 2-gram setup with 34 features, which is the one that yielded the highest accuracy. In our case, the time required for the same experiment, including the n-fold validation, is between 10-20 seconds, and close to one minute including the feature computation of all the SLDs of the dataset, without considering parallelisation. Therefore, the trade-off between accuracy (which in our case is just 2\% lower) and computational time (i.e. two orders of magnitude faster) of our method is clearly outperforming the method presented in \cite{SELVI2019156}.

\subsection{Multiclass Classification}
Considering related works that have studied the multiclass classification (see \autoref{tab:multi_stats}), a fine-grained comparison of processing performance was not possible since the achieved performance was not systematically reported and in some cases not stated at all. Moreover, such performance highly varies depending on the exact hardware configuration, and thus, the exact replication of each of the environments used is challenging. For instance, the authors of~\cite{stefanotracking} report only that experiments required time in the order of minutes.
In the case of~\cite{Anderson2016}, authors stated that their model required expensive training periods of 14 hours (considering 300 epochs and Alexa subsampling), as well as a classification time of 7 minutes for approximately 13k samples (i.e. 0.03 seconds per sample) in a GPU-based setup.
In work presented in~\cite{ALAEIYAN2020}, the authors reported a feature computation time of around 217 minutes for 252,757 samples, which implies approximately 0.05 seconds per sample. Moreover, the authors also specified a classification time of around 60 minutes, which translates into 0.014 seconds on average to classify a sample.

%In all cases,  considering our system configurations and the computational costs of feature computation, training, and classification,
In sum, our approach outperformed related work in the multiclass experiment in time requirements by one order of magnitude for both the feature computation as well as in the prediction. %, without using parallelisation.
This means that our method is suitable for real-time AGD detection and classification, even in environments with very high traffic volumes.

\section{Discussion}
\label{sec:discussion}

In this section, we focus on a quantitative comparison of the most relevant DGA detectors of the related work, and we analyse the existing limitations in the literature.

\subsection{Quantitative Comparison}
The DGA research field has several open challenges; one of them is the continuous appearance of new families.
In contrast to other research fields that rely on standardised benchmarks (e.g., computer vision), DGA-based datasets need to be updated frequently to be able to prove the performance of the detection methods is of relevance in real scenarios.
Recently, Zago et al.~\cite{zago2020} created a balanced and structured dataset containing 38 families.
Nevertheless, although their approach is sound, it does not keep pace with the recent evolution of malware campaigns.
Solving this issue is hard, and although there are research efforts in adversarial classification (see \autoref{sec:background}), we argue that further research should focus on upgradeable versions of datasets and include version tracking.
In our dataset, we introduce a collection of 105 families, which better reflects the complexity and challenge of the current landscape.
Moreover, it is worth to note that obtaining enough samples of specific families can be a cumbersome and difficult task due to a number of reasons, such as the inner algorithmic structure of the DGA\footnote{E.g., it can create only a small set of samples due to the use of specific dictionaries.} or the fact that it has not been reverse engineered yet.
Therefore, we argue that obtaining a perfectly balanced dataset that contains all possible DGA families is extremely difficult, borderline impossible. In this regard, it should be stressed that due to the heterogeneity and non-replicability of some datasets as well as the range of different techniques applied in related work, a perfect and fair comparison is unfeasible.
\begin{table}[th]
	\centering
	\caption{A quantitative comparison of our work with the most relevant state-of-the-art approaches.}
	% Authors use diverse variants of neural networks such as deep (DNN), long-shot term (LSTM), recurrent (RNN), and convolutional (CNN). % IH: most of the shortcuts are well known and some were already pronouned in related work.
	% HMM stands for Hidden Markov Model. % HMM was already defined
	\label{tab:comparisonall}
	\scriptsize
	\rowcolors{2}{gray!25}{white}
\resizebox{\textwidth}{!}{%
	\begin{tabular}{cp{3.0cm}ccp{1.2in}p{1.3in}}
		\toprule
		\textbf{Ref.}  &  \textbf{Features} & \textbf{\# DGAs}& \textbf{AGD samples} &  \textbf{Method} & \textbf{Dataset Source} \\
		\midrule

		\cite{Anderson2016} & Lexical, entropy & 10 & 110,000& GAN, LSTM, RF & Manually crafted \\
		\cite{stefanotracking} & Lexical & 5 &1,153,516 & DBSCAN & SIE framework,  Exposure blacklist and other public implementations\\
		\cite{curtin2018detecting} & WHOIS, lexical, smashword score  & 41 &1,280,000 & RNN & DGArchive and Several GitHub repositories  \\
		\cite{koh2018inline} & Lexical  &   4 & 4,000& ELMo  & Several GitHub repositories\\
		\cite{yu2017inline}& Time-based, query response, domain name    &  19 & 4,739,563&  LSTM / CNN  & Farsight security / DGArchive   \\
		\cite{18}& Domain name    &    37  & 169,831& LSTM  &  Bambenek   \\
		\cite{lison2017automatic} &Domain name    &  58 & 2,900,000& RNN  &  DGArchive, Bambenek   \\
		\cite{3}& Lexical     &  N/A  & 1,000,000&  CCN/RNN  &  Bambenek   \\
		\cite{mac2017dga}&  Entropy, lexical      &37 & 81,490&  Several ML methods  &Bambenek   \\
		\cite{choudhary2018algorithmically}& Lexical      &  19  &34,264,306&  Random Forest and DNN  &  Bambenek, DGArchive  \\
		\cite{li2019machine}& Lexical, query response     &  5  &160,000 &  Several ML methods   &  Bambenek   \\
		\cite{jyothsna2018detecting}& Lexical     &  19  &245,872&  DNN  &  Bambenek, Netlab 360   \\
		\cite{chen2018towards}& Domain name  &  60  &1,687,806&  LSTM  &  Bambenek, Netlab 360   \\
		\cite{sivaguru2018evaluation} & Time-based and domain name    &   15  &551,086&  Several Binary Classificators  & Real traffic, Bambenek  \\
		\cite{attardi2018bidirectional} &Domain name  &    19  & 135,056& LSTM, BLSTM  & Bambenek and Netlab 360  \\
		\cite{Zago2019}& Entropy, lexical, similarity    &  17 & 16,000& Several ML methods  &  Netlab 360, DGArchive, DNS-BH  \\
		\cite{Bharathi2019678}  & Domain name   &   19  & 245,872&  LSTM, BLSTM  & Bambenek, Netlab 360   \\
		\cite{khehra2018botscoop} & Entropy, lexical  &  5 &272,209 &  CNN/RNN  & Stratosphere dataset~\cite{stratosphereips}  \\
		\cite{zago2020} & Lexical & 38& 30,799,449 &Several ML methods & UMUDGA \\
		\cite{ANAND20201129} & Lexical & 19 & 25,300& Several ML methods &  Netlab 360 \\
		\cite{9072447} & Domain name  & 20 & 100,000& BLSTM, HDNN & Fu et al.~\cite{7852496} \\
		\cite{ALAEIYAN2020} & Lexical, pronounceability  & 30 &252,757 & Genetic algorithm and RF & Bader~\cite{badergithub} \\
		\cite{ALMASHHADANI2020101787} &Entropy, randomness, lexical &  20 & 208,190& Several ML methods  &  DGArchive, Bambenek  \\
		\cite{SELVI2019156} &Entropy, lexical &  26 & 252,757& RF  & Bader~\cite{badergithub}  \\
		\specialcell{Our Approach} & Entropy, lexical, gibberish  & 105 &95,325,598 & RF & HYDRA Dataset~\cite{hydra_dataset} \\
		\bottomrule
	\end{tabular}
	}
\end{table}
As it can be observed in \autoref{tab:comparisonall}, the results of our work were achieved using the biggest dataset in terms of the number of DGA families (and samples). This compares favourably and leads to a much more challenging classification task when compared to related work.
Due to the continuous evolution of malware, a high number of supported DGA families (including the most recent ones) is a critical capability required of any successful DGA detector that can be deployed in real scenarios.
Note that data sources of related works in the table share some common repositories such as Bambenek, DGArchive, and NetLab 360.
However, only a few authors used reverse-engineered DGAs to populate their datasets, as seen in \autoref{tab:comparisonall} and described in \autoref{sec:data}.
With regard to the detection method used, LSTM is the most prevalent, followed by ML classifiers, from which RF is known to report the best classification performance. Regarding the features, both lexical (e.g., ratios of letters, n-grams, words) and entropy-based ones seem to occur most widely.
%

%Moreover, it is important to note that the analysis of the outcomes in a benign dataset is essential as one could otherwise lower some feature thresholds (e.g., gibberish) and hence boost the accuracy quite significantly by applying more strict policies. However, due to the use of abbreviations, slang, etc. benign domains end up containing gibberish-like domains.

The methods that use side-information (e.g., WHOIS, timing, etc.) cannot prevent compromised hosts from contacting the C\&C server, and thus bring an additional cost in terms of time which makes them prohibitively slow for real-time detection and incident response.
Undeniably, caching and whitelisting can significantly reduce such a cost; however, this is expected to occur every time the host has to connect with a new domain or a DGA has a new seed, which is unrealistic in most scenarios.

\subsection{Fair Comparison and Evaluation of Reproducibility}
As previously reported in \autoref{sec:classification}, the comparison of any two approaches should be made under the same contextual settings (i.e., benchmarks and performance metrics); otherwise, the interpretation of the results might be biased and unduly favour an approach with less challenging settings.
To analyse the quality and the methodologies used in related research, we reviewed the works from \autoref{tab:comparisonall} in terms of reproducibility and presentation of the outcomes.
In detail, we verified whether the authors explicitly reported their evaluation methodology, their dataset collection procedure (for reproducibility purposes), and sufficient details about their outcomes (for a fair comparison). % (i.e., aggregated or per family). % so that proper comparison can be performed
The results of this effort are shown in \autoref{tab:stateofpractice}, where we observe that there are some serious methodological issues mainly due to the lack of experimental setup description, biased performance measures (e.g not reporting widely used metrics to enable fair comparison) being used and/or extremely imbalanced datasets (e.g. not enough samples for unbiased training), and the aggregation of classification results by averaging, while not reporting information about some poorly performing families. Note that \autoref{tab:stateofpractice} is not intended to criticise the related works, on the contrary, it aims to establishing a common ground to improve the transparency and contributions of the literature.
\begin{table}[th]
	\rowcolors{2}{gray!25}{white}
	\centering
	\footnotesize

	\caption{Methodological limitations of related work.}
	\label{tab:stateofpractice}
	\begin{tabular}{p{.40\linewidth}p{.50\linewidth}}
		\toprule \textbf{Methodological Limitations} & \textbf{References}\\
		\midrule
		% Unbiased DB &~\cite{Anderson2016, stefanotracking,koh2018inline,choudhary2018algorithmically,li2019machine,sivaguru2018evaluation}~\cite{attardi2018bidirectional,Bharathi2019678,zago2020,ALMASHHADANI2020101787,9072447,8936543} \\
		Not reported samples, extremely imbalanced benchmarks or lack of robust performance measures &\cite{yu2017inline,curtin2018detecting,18,lison2017automatic,3,mac2017dga,chen2018towards,ANAND20201129,chen2018towards,Zago2019,khehra2018botscoop}\\
		Aggregated classification outcomes  &\cite{18,lison2017automatic,3,choudhary2018algorithmically,jyothsna2018detecting,chen2018towards,chen2018towards,sivaguru2018evaluation,ANAND20201129,attardi2018bidirectional,Zago2019,Bharathi2019678,khehra2018botscoop,zago2020,ALAEIYAN2020,ALMASHHADANI2020101787,9072447,SELVI2019156}\\
		\bottomrule
	\end{tabular}
\end{table}

%In terms of evaluation reproducibility, a comparison among methods should only be performed using the same benchmarks and performance metrics.
%This is not possible in several cases, mainly due to the lack of description of the benchmarks used. % IH: repeated
Moreover, when the reported results are aggregated (i.e. the outcomes are not reported per class but as an overall aggregate), the unbalanced nature of the dataset, as well as the fact of hiding the classification performance per family hinders the objective interpretation of the results (e.g., a very small sample set could not reflect the characteristics of a family, and specific sampling ratios of benign to malicious domains might result in statistically biased outcomes).
This, in turn, translates into approaches that might obtain highly accurate results for some families, while they are unable to detect other families; however, the occurrence of this phenomenon cannot be discerned from the reported results, and thus a fair comparison cannot be made.

The same benchmark settings in the case of multiclass classification are even more critical to allow for a fair comparison since the more families used the more difficult is to classify them, especially taking into account their random nature. Moreover, since several DGAs can create the same pattern, the more samples collected the more possibilities of overlapping the domain names~\cite{ALAEIYAN2020,zago2020,patsakis2019exploiting,mac2017dga}, which increases chances for misclassification thus making the problem more challenging.
Nevertheless, in the case of binary classification with a statistically sound methodology, highly accurate detection of underrepresented families indicates the robustness of the selected features, thus showcasing the high performance of the detection method even in extreme cases.

\subsection{Sound Evaluation Methodology}
We argue that an objective comparison of the results among different approaches is possible only through a sound evaluation methodology.
To do so, it is imperative to include the reporting of the performance over data with the same ratio of AGD samples to benign domains.
Moreover, such experiments should be repeated several times with different sample sets (e.g., 100 times in our case and using 10-fold cross-validation in each iteration), since methodologies using a single run of n-fold cross-validation only shuffles samples within the selected sample set.
For instance, one could perform a 1:1 ratio classification between Alexa and a malicious family with 50 samples. In this setup, which is frequently adopted by ML practitioners in this field, only 50 samples of Alexa would be selected and shuffled in the cross-validation.
Therefore, the rest of the Alexa samples will not be used unless the experiments are repeated with different samples to produce a statistically sound outcome.
%Note that using the whole benign dataset, if it is several orders of magnitude larger and contains samples that could be misclassified\ih{not clear}, would affect the classification outcomes (i.e. a very small subset of the benign dataset could be misclassified, yet such impact would be proportional to the size of the malicious family represented).
In this regard, repeating the experiments and selecting a different set of samples in each iteration provides a better representation of the characteristics of each family and hence a much more realistic accuracy. % and avoids class imbalance issues.
In this setup, low values of the standard deviation in the results indicate the desirable stability and robustness of the method. Unfortunately, this methodology is rarely adopted in the field.

\subsection{Ratios of Malicious to Benign Samples}
The well-known imbalance problem~\cite{4717268} argues that there are much less malicious events than benign ones when performing, e.g., traffic analysis and intrusion detection in real scenarios~\cite{shabtai2012detecting,quing2020,liu2018imbalanced}. Nevertheless, in the area of DGA analysis, there is no wide consensus on the common ratio of benign to malicious domains that one can commonly find in real-world settings. This is due to some DGAs generating only a few domains per day, while others might create them in the hundreds or thousands.
Therefore, we made a conscious effort to evaluate the performance of our approach under malicious to benign ratios different than 1:1, and hence we explored ratios of 1:10 and 1:100 as well.
During our experiments, if a malicious family had more samples than the size of the benign dataset, these were randomly under-sampled, to obtain the desired ratios.
\autoref{tab:binary_stats_samples} shows the results obtained by using the $F_1$ measure and its standard deviation across 100 repetitions of the 10-fold cross-validation, selecting different samples in each iteration.

\begin{table}[t]
	\setlength{\tabcolsep}{3pt}
	\def\arraystretch{1.1}
	\centering
	\caption{Binary classification outcomes with different ratios of malicious to benign domains (we always assume a higher number of benign domains in the ratios).}
	%	The lower the standard deviation, the more statistically sound the results. % IH: too detailed for a caption
	\label{tab:binary_stats_samples}
	\scriptsize
	\resizebox{\textwidth}{!}{%
	\begin{tabular}{>{\bfseries}lcccc>{\bfseries \columncolor{gray!25}}l>{\columncolor{gray!25}}c>{\columncolor{gray!25}}c>{\columncolor{gray!25}}c>{\columncolor{gray!25}}c>{\bfseries}lcccc}
		\hline
		& \multicolumn{2}{c}{\textbf{Ratio 1:10}}  & \multicolumn{2}{c}{\textbf{Ratio 1:100}} & & \multicolumn{2}{c}{\cellcolor{gray!25}\textbf{Ratio 1:10}}  & \multicolumn{2}{c}{\cellcolor{gray!25}\textbf{Ratio 1:100}} & & \multicolumn{2}{c}{\textbf{Ratio 1:10}}  & \multicolumn{2}{c}{\textbf{Ratio 1:100}} \\
		\textbf{Class}                 & \textbf{$F_1$}  & \textbf{$\sigma$} & \textbf{$F_1$}      & \textbf{$\sigma$}     & \textbf{Class}               & \textbf{$F_1$}  & \textbf{$\sigma$} & \textbf{$F_1$}      & \textbf{$\sigma$}   & \textbf{Class}                  & \textbf{$F_1$}  & \textbf{$\sigma$} & \textbf{$F_1$}      & \textbf{$\sigma$}     \\
		\hline
bamital                                   & 99.98 & 0.03            & 99.97 & 0.03 & gspy                           & 100   & 0               & 100   & 0               & qakbot                 & 99.88 & 0.03 & 99.87 & 0.03            \\
banjori                                   & 99.60  & 0.13            & 99.51 & 0.18 & hesperbot & 99.90  & 0.05            & 99.90  & 0.05            & qhost                  & 100   & 0    & 100   & 0               \\
bedep                                     & 99.93 & 0.03            & 99.93 & 0.03 & infy                           & 99.92 & 0.03            & 99.93 & 0.08            & qsnatch                & 99.07 & 0.36 & 98.82 & 0.14            \\
beebone                                   & 99.53 & 0.40             & 99.53 & 0.40  & khaos                          & 93.33 & 0.69            & 92.96 & 0.83            & ramdo                  & 99.98 & 0.03 & 99.98 & 0.03            \\
bigviktor                                 & 93.56 & 0.58            & 93.26 & 0.28 & kingminer                      & 98.80  & \textless{}0.01 & 98.80  & \textless{}0.01 & ramnit                 & 99.90  & 0.05 & 99.93 & 0.03            \\
blackhole                                 & 100   & 0               & 99.98 & 0.04 & locky                          & 99.63 & 0.06            & 99.77 & 0.12            & ranbyus                & 99.93 & 0.06 & 99.95 & 0.05            \\
\specialcell{bobax/\\~~/kraken\\~~/oderoor} & 99.82 & 0.03            & 99.68 & 0.12 & madmax                         & 99.90  & 0.05            & 99.93 & 0.06            & redyms                 & 100   & 0    & 100   & 0               \\
ccleaner             & 100   & 0               & 99.98 & 0.03 & makloader                      & 99.94 & 0.11            & 100   & 0               & rovnix                 & 100   & 0    & 99.98 & 0.03            \\
chinad                                    & 99.97 & 0.03            & 100   & 0    & matsnu                         & 91.86 & 0.67            & 91.58 & 0.46            & shifu                  & 99.68 & 0.10  & 99.70  & 0.05            \\
chir                                      & 100   & 0               & 100   & 0    & mirai                          & 99.95 & 0.05            & 99.90  & 0.05            & \specialcell{shiotob/\\~~/urlzone\\~~/bebloh} & 99.95 & 0.05 & 99.98 & 0.03            \\
conficker                                 & 99.45 & 0.05            & 99.23 & 0.13 & modpack                        & 100   & 0               & 100   & 0               & simda                  & 99.03 & 0.10  & 99.09 & 0.14            \\
corebot                                   & 99.97 & 0.03            & 99.97 & 0.03 & \specialcell{monerodownloader}               & 100   & 0               & 100   & 0               & sisron                 & 100   & 0    & 100   & 0               \\
cryptolocker                              & 99.95 & 0.05            & 99.95 & 0.05 & monerominer                    & 100   & 0               & 100   & 0               & sphinx                 & 100   & 0    & 99.97 & 0.06            \\
cryptowall                                & 99.87 & 0.06            & 99.95 & 0.05 & murofet                        & 100   & 0               & 99.95 & \textless{}0.01 & suppobox               & 92.39 & 0.37 & 91.92 & 0.59            \\
darkshell                                 & 100   & 0               & 99.58 & 0.73 & murofetweekly                  & 100   & 0               & 100   & 0               & sutra                  & 99.95 & 0.05 & 99.97 & 0.06            \\
deception            & 95.02 & 0.29            & 94.69 & 0.44 & mydoom                         & 99.73 & 0.03            & 99.75 & 0.05            & symmi                  & 96.55 & 0.48 & 96.55 & 0.48            \\
deception2           & 92.85 & 0.86            & 92.49 & 0.46 & necurs                         & 99.85 & 0.09            & 99.87 & 0.03            & szribi                 & 99.70  & 0.09 & 99.60  & 0.17            \\
diamondfox                                & 98.61 & 0.10             & 98.67 & 0.10  & nymaim                         & 99.46 & 0.12            & 99.38 & 0.10             & tempedreve             & 99.62 & 0.06 & 99.7  & 0.10             \\
dircrypt             & 99.85 & \textless{}0.01 & 99.92 & 0.03 & nymaim2                        & 94.53 & 0.58            & 94.37 & 0.41            & tinba                  & 99.92 & 0.03 & 99.95 & 0.05            \\
dmsniff                                   & 98.06 & 0.43            & 98.32 & 0.39 & omexo                          & 99.55  & 0.69             & 99.60  & 0.70             & tinynuke               & 100   & 0    & 100   & 0               \\
dnschanger                                & 99.95 & 0.05            & 99.88 & 0.03 & padcrypt                       & 99.98 & 0.03            & 100   & 0               & tofsee                 & 99.90  & 0.05 & 99.80  & 0               \\
dromedan             & 99.9  & 0.10             & 99.93 & 0.03 & pandabanker                    & 99.98 & 0.03            & 99.97 & 0.06            & torpig                 & 99.48 & 0.13 & 99.62 & 0.03            \\
dyre                                      & 100   & 0               & 100   & 0    & pitou                          & 99.87 & 0.03            & 99.80  & 0.13            & tsifiri                & 99.16 & 0.83 & 99.72 & 0.49            \\
ebury                                     & 99.98 & 0.03            & 100   & 0    & pizd                           & 96.90  & 0.09            & 96.61 & 0.25            & ud2                    & 100   & 0    & 99.97 & 0.06            \\
ekforward                                 & 99.92 & 0.03            & 99.93 & 0.03 & post                           & 100   & 0               & 100   & 0               & ud3                    & 100   & 0    & 100   & 0               \\
emotet                                    & 99.98 & 0.03            & 99.98 & 0.03 & proslikefan                    & 99.75 & 0.09            & 99.51 & 0.15            & ud4                    & 98.30  & 0.43 & 98.30  & 0.43            \\
enviserv                                  & 99.97 & 0.06            & 100   & 0    & pushdo                         & 98.46 & 0.06            & 98.28 & 0.16            & vawtrak                & 98.77 & 0.19 & 98.54 & 0.36            \\
feodo                                     & 99.93 & 0.12            & 99.93 & 0.12 & pushdotid                      & 99.72 & 0.08            & 99.73 & 0.06            & vidro                  & 99.83 & 0.08 & 99.88 & 0.03            \\
fobber                                    & 99.93 & 0.03            & 99.88 & 0.03 & pykspa                         & 99.67 & 0.08            & 99.67 & 0.18            & vidrotid               & 97.82 & 0.28 & 97.98 & \textless{}0.01 \\
fobber\_v1                                & 100   & 0               & 100   & 0    & pykspa\_v1                     & 99.58 & 0.03            & 99.63 & 0.08            & virut                  & 99.80  & 0.09 & 99.88 & 0.03            \\
fobber\_v2                                & 100   & 0               & 100   & 0    & pykspa\_v2\_fake               & 99.65 & 0.05            & 99.51 & 0.08            & volatilecedar          & 99.90  & 0.10  & 99.93 & 0.12            \\
gameover                                  & 100   & 0               & 99.98 & 0.03 & pykspa\_v2\_real               & 99.55 & 0.09            & 99.58 & 0.03            & wd                     & 100   & 0    & 100   & 0               \\
geodo                                     & 100   & 0               & 100   & 0    & pykspa2                        & 99.52 & 0.04            & 99.56 & \textless{}0.01 & xshellghost            & 99.93 & 0.03 & 99.85 & 0.05            \\
gozi                 & 93.04 & 0.19            & 92.68 & 0.52 & pykspa2s                       & 98.29 & 0.15            & 98.20  & \textless{}0.01 & xxhex                  & 100   & 0    & 99.95 & 0.09            \\
goznym                                    & 99.54 & 0.08            & 99.49 & 0.08 & qadars                         & 99.93 & 0.03            & 99.93 & 0.03            & zloader                & 100   & 0    & 100   & 0              \\
		\bottomrule
	\end{tabular}
	}
\end{table}

When we compare the outcomes obtained across all ratios (i.e. 1:1, 1:10, and 1:100), we can observe that dictionary-based and adversarial families obtain slightly worse accuracy when the malicious to benign ratio is increased. The latter occurs because these DGAs create domains that have structural similarities with Alexa domains, which increases the difficulty of the classification task due to the overlapping features.
Nevertheless, as it can be observed in the rest of cases, the variance of the outcomes according to each sampling ratios is minimal, which denotes stable results. This means that our approach and its features represent homogeneously benign domains, and thus, they are able to accurately distinguish them from malicious ones, regardless of the sample ratio. This showcases the quality of the feature selection as well as the statistical confidence of the classification.

A proper methodology should also be considered when using automated approaches for machine learning, such as H2O\footnote{\url{https://www.h2o.ai/}}, auto-sklearn \footnote{\url{https://github.com/automl/auto-sklearn}}, AutoKeras\footnote{\url{https://github.com/keras-team/autokeras?spm=a2c65.11461447.0.0.68b37903yEmaw3}} etc.
Such libraries may hyper-optimise parameters for many methods and generate a model which maximises, for instance, the $F_1$ score. We argue that this \textit{unique win} should not be considered as the best method since, as discussed above, this solution has to be weighed along with the efficiency of the rest of the models over the same family, and considering several repetitions, only in this way providing statistical soundness.

\section{Conclusion}
\label{sec:conclusions}
%\ih{it seems too long and verbose to me}
Nowadays, modern malware has evolved into highly sophisticated software, which can be used to infect millions of devices. This enables hard-to-detect and resilient malware campaigns, which have turned cybercrime into a profitable ``business''. To enable faster and more accurate botnet detection, and to speed-up take-down operations, a new DGA detection method using machine learning is presented.
In essence, our method stands out from the rest in terms of accuracy and performance because we use more comprehensive features and a broader and more representative dataset. We only identified a case in which our outcomes were slightly below these obtained by other methods, yet the time required was between one and two orders of magnitude lower in our case. The relevance of our features is manifested in three ways. First, it achieves an almost optimal detection rate in the binary classification problem for the broadest possible set of DGA families. Second, our features allow us to outperform the current state-of-the-art also in multiclass classification, using the same datasets presented in other works. Finally, our approach was able to detect adversarially designed DGAS, including the experiments in which our system was not trained to detect such families (i.e. assuming no previous knowledge).

Additionally, our methodology is more rigorous than most seen in the field to date, avoiding common pitfalls in the literature that focus on DGAs with many non-obvious constraints. Setting aside feature extraction, our work highlights the inherent biases of datasets and methodologies in previous literature that report many close to perfect results; however, these results may be true for only a very limited and unrepresentative number of DGA families. Notably, we stress the methodological errors in the use of machine learning with, e.g., the use of very few samples and in some cases aggregated classification outcomes preventing a clear comparison.
In this regard, a dataset with more than 95 million AGDs is constructed and shared, providing the extracted features to the community. While this facilitates the reproducibility of our results, we also allow fellow researchers to use a significantly richer baseline dataset, both in terms of number of families and samples.

In future work, we aim to enhance our semantic classification by using other training sources in order to increase the accuracy of both English and non-English domain names. Moreover, we will explore wordlist-based DGA detection in more depth by using probabilistic approaches based on word repetition and similar features. Finally, we will study the impact of dimensionality reduction techniques in our dataset.

\section*{Acknowledgement}
This work was supported by the European Commission under the Horizon 2020 Programme (H2020), as part of the projects CyberSec4Europe (\url{https://www.cybersec4europe.eu}) (Grant Agreement no. 830929), \textit{LOCARD} (\url{https://locard.eu}) (Grant Agreement no. 832735) and \textit{YAKSHA}, (\url{https://project-yaksha.eu/project/}) (Grant Agreement no 780498).
Also, this work was supported by the H2020 ECSEL project VALU3S (876852) and the internal project of Brno University of Technology (FIT-S-20-6427).
The Titan Xp used for this research was generously donated by NVIDIA Corporation.

The content of this article does not reflect the official opinion of the European Union. Responsibility for the information and views expressed therein lies entirely with the authors.

% \bibliographystyle{plain}
% \bibliography{refs}

\begin{thebibliography}{10}

\bibitem{abakumovgithub}
Andrey Abakumov.
\newblock Dga repository.
\newblock \url{https://github.com/andrewaeva/DGA}, 2020.

\bibitem{ALAEIYAN2020}
Mohammadhadi Alaeiyan, Saeed Parsa, Vinod P., and Mauro Conti.
\newblock Detection of algorithmically-generated domains: An adversarial
  machine learning approach.
\newblock {\em Computer Communications}, 2020.

\bibitem{ALMASHHADANI2020101787}
Ahmad~O. Almashhadani, Mustafa Kaiiali, Domhnall Carlin, and Sakir Sezer.
\newblock Maldomdetector: A system for detecting algorithmically generated
  domain names with machine learning.
\newblock {\em Computers \& Security}, 93:101787, 2020.

\bibitem{ANAND20201129}
P.~Mohan Anand, T.~Gireesh Kumar, and P.V.~Sai Charan.
\newblock An ensemble approach for algorithmically generated domain name
  detection using statistical and lexical analysis.
\newblock {\em Procedia Computer Science}, 171:1129 -- 1136, 2020.
\newblock Third International Conference on Computing and Network
  Communications (CoCoNet'19).

\bibitem{Anderson2016}
Hyrum~S. Anderson, Jonathan Woodbridge, and Bobby Filar.
\newblock {DeepDGA}: Adversarially-tuned domain generation and detection.
\newblock In {\em Proceedings of the 2016 ACM Workshop on Artificial
  Intelligence and Security}, AISec '16, pages 13--21, New York, NY, USA, 2016.
  ACM.

\bibitem{ipfsstorm}
{Anomali Labs}.
\newblock Interplanetary storm.
\newblock
  \url{https://www.anomali.com/blog/the-interplanetary-storm-new-malware-in-wild-using-interplanetary-file-systems-ipfs-p2p-network},
  2019.

\bibitem{1}
Manos Antonakakis et~al.
\newblock From throw-away traffic to bots: detecting the rise of {DGA}-based
  malware.
\newblock In {\em Proceedings of the 21st USENIX conference on Security
  symposium}, pages 24--24. USENIX Association, 2012.

\bibitem{203628}
Manos Antonakakis et~al.
\newblock Understanding the mirai botnet.
\newblock In {\em 26th {USENIX} Security Symposium ({USENIX} Security 17)},
  pages 1093--1110, Vancouver, BC, 2017. {USENIX} Association.

\bibitem{attardi2018bidirectional}
Giuseppe Attardi and Daniele Sartiano.
\newblock Bidirectional lstm models for dga classification.
\newblock In {\em International Symposium on Security in Computing and
  Communication}, pages 687--694. Springer, 2018.

\bibitem{Aviv2011}
Adam~J. Aviv and Andreas Haeberlen.
\newblock Challenges in experimenting with botnet detection systems.
\newblock In {\em Proceedings of the 4th Conference on Cyber Security
  Experimentation and Test}, CSET'11, pages 6--6, Berkeley, CA, USA, 2011.
  USENIX Association.

\bibitem{johannesbader}
Johannes Bader.
\newblock The {DGA} of pykspa ``you skype version is old''.
\newblock \url{https://www.johannesbader.ch/2015/03/the-dga-of-pykspa/}, 2015.

\bibitem{badergithub}
Johannes Bader.
\newblock Domain generation algorithms (dgas) of malware reimplemented in
  python.
\newblock \url{https://github.com/baderj/domain_generation_algorithms}, 2020.

\bibitem{barandela2003new}
Ricardo Barandela, Rosa~Maria Valdovinos, and Jos{\'e}~Salvador S{\'a}nchez.
\newblock New applications of ensembles of classifiers.
\newblock {\em Pattern Analysis \& Applications}, 6(3):245--256, 2003.

\bibitem{berman2019dga}
Daniel~S Berman.
\newblock Dga capsnet: 1d application of capsule networks to dga detection.
\newblock {\em Information}, 10(5):157, 2019.

\bibitem{Bharathi2019678}
B.~Bharathi and J.~Bhuvana.
\newblock Domain name detection and classification using deep neural networks.
\newblock {\em Communications in Computer and Information Science},
  969:678--686, 2019.

\bibitem{braden1989rfc1123}
Robert Braden.
\newblock Rfc1123: Requirements for internet hosts-application and support,
  1989.

\bibitem{breiman1996bagging}
Leo Breiman.
\newblock Bagging predictors.
\newblock {\em Machine learning}, 24(2):123--140, 1996.

\bibitem{hydra_dataset}
Fran Casino, Nikolaos Lykousas, Ivan Homoliak, Constantinos Patsakis, and Julio
  Hernandez-Castro.
\newblock Hydra dataset, July 2020.

\bibitem{chaignongithub}
Paul Chaignon.
\newblock Dga collection.
\newblock \url{https://github.com/pchaigno/dga-collection}, 2020.

\bibitem{chen2018towards}
Yang Chen, Shuai Zhang, Jing Liu, and Bo~Li.
\newblock Towards a deep learning approach for detecting malicious domains.
\newblock In {\em 2018 IEEE International Conference on Smart Cloud
  (SmartCloud)}, pages 190--195. IEEE, 2018.

\bibitem{choudhary2018algorithmically}
Chhaya Choudhary et~al.
\newblock Algorithmically generated domain detection and malware family
  classification.
\newblock In {\em International Symposium on Security in Computing and
  Communication}, pages 640--655. Springer, 2018.

\bibitem{curtin2018detecting}
Ryan~R Curtin, Andrew~B Gardner, Slawomir Grzonkowski, Alexey Kleymenov, and
  Alejandro Mosquera.
\newblock Detecting dga domains with recurrent neural networks and side
  information.
\newblock In {\em Proceedings of the 14th International Conference on
  Availability, Reliability and Security}, pages 1--10, 2019.

\bibitem{dietterich2000ensemble}
Thomas~G Dietterich.
\newblock Ensemble methods in machine learning.
\newblock In {\em International workshop on multiple classifier systems}, pages
  1--15. Springer, 2000.

\bibitem{7852496}
Y.~{Fu}, L.~{Yu}, O.~{Hambolu}, I.~{Ozcelik}, B.~{Husain}, J.~{Sun},
  K.~{Sapra}, D.~{Du}, C.~T. {Beasley}, and R.~R. {Brooks}.
\newblock Stealthy domain generation algorithms.
\newblock {\em IEEE Transactions on Information Forensics and Security},
  12(6):1430--1443, June 2017.

\bibitem{homoliak2014nba}
Ivan Homoliak, Daniel Ovsonka, Matej Gregr, and Petr Hanacek.
\newblock Nba of obfuscated network vulnerabilities' exploitation hidden into
  https traffic.
\newblock In {\em The 9th International Conference for Internet Technology and
  Secured Transactions (ICITST-2014)}, pages 310--317. IEEE, 2014.

\bibitem{5762763}
N.~Jiang, J.~Cao, Y.~Jin, L.~E. Li, and Z.~Zhang.
\newblock Identifying suspicious activities through dns failure graph analysis.
\newblock In {\em The 18th IEEE International Conference on Network Protocols},
  pages 144--153, Oct 2010.

\bibitem{jyothsna2018detecting}
PV~Jyothsna, Greeshma Prabha, KK~Shahina, and Anu Vazhayil.
\newblock Detecting dga using deep neural networks (dnns).
\newblock In {\em International Symposium on Security in Computing and
  Communication}, pages 695--706. Springer, 2018.

\bibitem{khehra2018botscoop}
Gulbadan Khehra and Sanjeev Sofat.
\newblock Botscoop: Scalable detection of dga based botnets using dns traffic.
\newblock In {\em 2018 9th International Conference on Computing, Communication
  and Networking Technologies (ICCCNT)}, pages 1--6. IEEE, 2018.

\bibitem{klensin2010internationalized}
J~Klensin.
\newblock Internationalized domain names in applications (idna): Protocol.
\newblock Technical report, RFC 5891, August, 2010.

\bibitem{koh2018inline}
Joewie~J Koh and Barton Rhodes.
\newblock Inline detection of domain generation algorithms with
  context-sensitive word embeddings.
\newblock In {\em 2018 IEEE International Conference on Big Data (Big Data)},
  pages 2966--2971. IEEE, 2018.

\bibitem{kuncheva2014combining}
Ludmila~I Kuncheva.
\newblock {\em Combining pattern classifiers: methods and algorithms}.
\newblock John Wiley \& Sons, 2014.

\bibitem{li2019machine}
Yi~Li, Kaiqi Xiong, Tommy Chin, and Chengbin Hu.
\newblock A machine learning framework for domain generation algorithm
  (dga)-based malware detection.
\newblock {\em IEEE Access}, 2019.

\bibitem{lison2017automatic}
Pierre Lison and Vasileios Mavroeidis.
\newblock Automatic detection of malware-generated domains with recurrent
  neural models.
\newblock {\em arXiv preprint arXiv:1709.07102}, 2017.

\bibitem{Liu2017}
Daiping Liu, Zhou Li, Kun Du, Haining Wang, Baojun Liu, and Haixin Duan.
\newblock Don't let one rotten apple spoil the whole barrel: Towards automated
  detection of shadowed domains.
\newblock In {\em Proceedings of the 2017 ACM SIGSAC Conference on Computer and
  Communications Security}, CCS '17, pages 537--552, New York, NY, USA, 2017.
  ACM.

\bibitem{4717268}
X.~{Liu}, J.~{Wu}, and Z.~{Zhou}.
\newblock Exploratory undersampling for class-imbalance learning.
\newblock {\em IEEE Transactions on Systems, Man, and Cybernetics, Part B
  (Cybernetics)}, 39(2):539--550, 2009.

\bibitem{liu2018imbalanced}
Zhenyan Liu, Yifei Zeng, Pengfei Zhang, Jingfeng Xue, Ji~Zhang, and Jiangtao
  Liu.
\newblock An imbalanced malicious domains detection method based on passive dns
  traffic analysis.
\newblock {\em Security and Communication Networks}, 2018, 2018.

\bibitem{mac2017dga}
Hieu Mac, Duc Tran, Van Tong, Linh~Giang Nguyen, and Hai~Anh Tran.
\newblock Dga botnet detection using supervised learning methods.
\newblock In {\em Proceedings of the Eighth International Symposium on
  Information and Communication Technology}, pages 211--218. ACM, 2017.

\bibitem{yadavgraph}
Pratyusa~K. Manadhata, Sandeep Yadav, Prasad Rao, and William Horne.
\newblock Detecting malicious domains via graph inference.
\newblock In Miros{\l}aw Kuty{\l}owski and Jaideep Vaidya, editors, {\em
  Computer Security - ESORICS 2014}, pages 1--18, Cham, 2014. Springer
  International Publishing.

\bibitem{nadji2013beheading}
Yacin Nadji, Manos Antonakakis, Roberto Perdisci, David Dagon, and Wenke Lee.
\newblock Beheading hydras: performing effective botnet takedowns.
\newblock In {\em Proceedings of the 2013 ACM SIGSAC conference on Computer \&
  communications security}, pages 121--132, 2013.

\bibitem{nadji2017still}
Yacin Nadji, Roberto Perdisci, and Manos Antonakakis.
\newblock Still beheading hydras: Botnet takedowns then and now.
\newblock {\em IEEE Transactions on Dependable and Secure Computing},
  14(5):535--549, 2017.

\bibitem{Patsakis2019}
Constantinos Patsakis and Fran Casino.
\newblock Hydras and ipfs: a decentralised playground for malware.
\newblock {\em International Journal of Information Security}, Jun 2019.

\bibitem{PATSAKIS2021102725}
Constantinos Patsakis and Fran Casino.
\newblock Exploiting statistical and structural features for the detection of
  domain generation algorithms.
\newblock {\em Journal of Information Security and Applications}, 58:102725,
  2021.

\bibitem{patsakis2019exploiting}
Constantinos Patsakis and Fran Casino.
\newblock Exploiting statistical and structural features for the detection of
  domain generation algorithms.
\newblock {\em Journal of Information Security and Applications}, 2021.

\bibitem{PATSAKIS2020101614}
Constantinos Patsakis, Fran Casino, and Vasilios Katos.
\newblock Encrypted and covert dns queries for botnets: Challenges and
  countermeasures.
\newblock {\em Computers \& Security}, 88:101614, 2020.

\bibitem{6175908}
R.~Perdisci, I.~Corona, and G.~Giacinto.
\newblock Early detection of malicious flux networks via large-scale passive
  dns traffic analysis.
\newblock {\em IEEE Transactions on Dependable and Secure Computing},
  9(5):714--726, Sept 2012.

\bibitem{197187}
Daniel Plohmann, Khaled Yakdan, Michael Klatt, Johannes Bader, and Elmar
  Gerhards-Padilla.
\newblock A comprehensive measurement study of domain generating malware.
\newblock In {\em 25th {USENIX} Security Symposium ({USENIX} Security 16)},
  pages 263--278, Austin, TX, 2016. {USENIX} Association.

\bibitem{stefanotracking}
Stefano Schiavoni, Federico Maggi, Lorenzo Cavallaro, and Stefano Zanero.
\newblock Phoenix: Dga-based botnet tracking and intelligence.
\newblock In Sven Dietrich, editor, {\em Detection of Intrusions and Malware,
  and Vulnerability Assessment}, pages 192--211, Cham, 2014. Springer
  International Publishing.

\bibitem{SELVI2019156}
Jose Selvi, Ricardo~J. Rodríguez, and Emilio Soria-Olivas.
\newblock Detection of algorithmically generated malicious domain names using
  masked n-grams.
\newblock {\em Expert Systems with Applications}, 124:156 -- 163, 2019.

\bibitem{shabtai2012detecting}
Asaf Shabtai, Robert Moskovitch, Clint Feher, Shlomi Dolev, and Yuval Elovici.
\newblock Detecting unknown malicious code by applying classification
  techniques on opcode patterns.
\newblock {\em Security Informatics}, 1(1):1, 2012.

\bibitem{silva2017improving}
Daniel Silva-Palacios, Cesar Ferri, and Mar{\'\i}a~Jos{\'e}
  Ram{\'\i}rez-Quintana.
\newblock Improving performance of multiclass classification by inducing class
  hierarchies.
\newblock {\em Procedia Computer Science}, 108:1692--1701, 2017.

\bibitem{SINGH201928}
Manmeet Singh, Maninder Singh, and Sanmeet Kaur.
\newblock Issues and challenges in dns based botnet detection: A survey.
\newblock {\em Computers \& Security}, 86:28 -- 52, 2019.

\bibitem{sivaguru2018evaluation}
Raaghavi Sivaguru et~al.
\newblock An evaluation of dga classifiers.
\newblock In {\em 2018 IEEE International Conference on Big Data (Big Data)},
  pages 5058--5067. IEEE, 2018.

\bibitem{7535098}
A.~K. Sood and S.~Zeadally.
\newblock A taxonomy of domain-generation algorithms.
\newblock {\em IEEE Security Privacy}, 14(4):46--53, July 2016.

\bibitem{Spooren2019}
Jan Spooren et~al.
\newblock Detection of algorithmically generated domain names used by botnets:
  A dual arms race.
\newblock In {\em Proceedings of the 34th ACM/SIGAPP Symposium on Applied
  Computing}, SAC '19, pages 1916--1923, New York, NY, USA, 2019. ACM.

\bibitem{stratosphereips}
{Stratosphere Labs}.
\newblock Stratosphere labs datasets overview.
\newblock \url{https://www.stratosphereips.org/datasets-overview}, 2020.

\bibitem{18}
Duc Tran, Hieu Mac, Van Tong, Hai~Anh Tran, and Linh~Giang Nguyen.
\newblock A lstm based framework for handling multiclass imbalance in dga
  botnet detection.
\newblock {\em Neurocomputing}, 275:2401--2413, 2018.

\bibitem{valentini2002ensembles}
Giorgio Valentini and Francesco Masulli.
\newblock Ensembles of learning machines.
\newblock In {\em Italian workshop on neural nets}, pages 3--20. Springer,
  2002.

\bibitem{Vinayakumar2019}
R.~Vinayakumar, K.~P. Soman, Prabaharan Poornachandran, S.~Akarsh, and Mohamed
  Elhoseny.
\newblock Improved dga domain names detection and categorization using deep
  learning architectures with classical machine learning algorithms.
\newblock In Aboul~Ella Hassanien and Mohamed Elhoseny, editors, {\em
  Cybersecurity and Secure Information Systems: Challenges and Solutions in
  Smart Environments}, pages 161--192. Springer International Publishing, Cham,
  2019.

\bibitem{quing2020}
Qing Wang et~al.
\newblock Malicious domain detection based on k-means and smote.
\newblock In {\em Computational Science -- ICCS 2020}, pages 468--481, Cham,
  2020. Springer International Publishing.

\bibitem{2}
Jonathan Woodbridge, Hyrum~S Anderson, Anjum Ahuja, and Daniel Grant.
\newblock Predicting domain generation algorithms with long short-term memory
  networks.
\newblock {\em arXiv preprint arXiv:1611.00791}, 2016.

\bibitem{XU201977}
Congyuan Xu, Jizhong Shen, and Xin Du.
\newblock Detection method of domain names generated by dgas based on semantic
  representation and deep neural network.
\newblock {\em Computers \& Security}, 85:77 -- 88, 2019.

\bibitem{6151233}
S.~Yadav, A.~K.~K. Reddy, A.~L.~N. Reddy, and S.~Ranjan.
\newblock Detecting algorithmically generated domain-flux attacks with {DNS}
  traffic analysis.
\newblock {\em IEEE/ACM Transactions on Networking}, 20(5):1663--1677, Oct
  2012.

\bibitem{yadav2012}
Sandeep Yadav and A.~L.~Narasimha Reddy.
\newblock Winning with {DNS} failures: Strategies for faster botnet detection.
\newblock In Muttukrishnan Rajarajan, Fred Piper, Haining Wang, and George
  Kesidis, editors, {\em Security and Privacy in Communication Networks}, pages
  446--459, Berlin, Heidelberg, 2012. Springer Berlin Heidelberg.

\bibitem{9072447}
L.~{Yang}, G.~{Liu}, Y.~{Dai}, J.~{Wang}, and J.~{Zhai}.
\newblock Detecting stealthy domain generation algorithms using heterogeneous
  deep neural network framework.
\newblock {\em IEEE Access}, 8:82876--82889, 2020.

\bibitem{yang2018novel}
Luhui Yang et~al.
\newblock A novel detection method for word-based dga.
\newblock In {\em International Conference on Cloud Computing and Security},
  pages 472--483. Springer, 2018.

\bibitem{yu2017inline}
Bin Yu, Daniel~L Gray, Jie Pan, Martine De~Cock, and Anderson~CA Nascimento.
\newblock Inline dga detection with deep networks.
\newblock In {\em 2017 IEEE International Conference on Data Mining Workshops
  (ICDMW)}, pages 683--692. IEEE, 2017.

\bibitem{3}
Bin Yu, Jie Pan, Jiaming Hu, Anderson Nascimento, and Martine De~Cock.
\newblock Character level based detection of dga domain names.
\newblock In {\em 2018 International Joint Conference on Neural Networks
  (IJCNN)}, pages 1--8. IEEE, 2018.

\bibitem{8936543}
X.~{Yun}, J.~{Huang}, Y.~{Wang}, T.~{Zang}, Y.~{Zhou}, and Y.~{Zhang}.
\newblock Khaos: An adversarial neural network dga with high anti-detection
  ability.
\newblock {\em IEEE Transactions on Information Forensics and Security},
  15:2225--2240, 2020.

\bibitem{Zago2019}
M.~Zago, M.~Gil~Pérez, and G.~Martínez~Pérez.
\newblock Scalable detection of botnets based on dga: Efficient feature
  discovery process in machine learning techniques.
\newblock {\em Soft Computing}, 2019.

\bibitem{zago2020}
Mattia Zago, Manuel~Gil Perez, and Gregorio~Martinez Perez.
\newblock Umudga: A dataset for profiling dga-based botnet.
\newblock {\em Computers \& Security}, 92:101719, 2020.

\bibitem{zander2007survey}
Sebastian Zander, Grenville Armitage, and Philip Branch.
\newblock A survey of covert channels and countermeasures in computer network
  protocols.
\newblock {\em IEEE Communications Surveys \& Tutorials}, 9(3):44--57, 2007.

\bibitem{8532279}
X.~{Zang}, J.~{Gong}, S.~{Mo}, A.~{Jakalan}, and D.~{Ding}.
\newblock Identifying fast-flux botnet with agd names at the upper dns
  hierarchy.
\newblock {\em IEEE Access}, 6:69713--69727, 2018.

\bibitem{7163279}
G.~Zhao, K.~Xu, L.~Xu, and B.~Wu.
\newblock Detecting {APT} malware infections based on malicious {DNS} and
  traffic analysis.
\newblock {\em IEEE Access}, 3:1132--1142, 2015.

\bibitem{Zhou2013DGABasedBD}
Yonglin Zhou, Qing-Shan Li, Qidi Miao, and Kangbin Yim.
\newblock {DGA}-based botnet detection using {DNS} traffic.
\newblock {\em J. Internet Serv. Inf. Secur.}, 3:116--123, 2013.

\end{thebibliography}

\appendix
\section{Feature Relevance}
\label{sec:feat_rel}
To showcase the relevance of the features used in our approach, the specific values of weights are provided in \autoref{tab:feature_w} in the case of our binary classification using the HYDRAS dataset (see \autoref{sec:classification}) and the multiclass classification performed with the dataset introduced in \cite{badergithub}. We computed the weights in different setups to highlight the fact that the importance of several features may vary according to the families analysed. The latter, which seems straightforward, is nevertheless worth showcasing given a specific subset of families, so that further insights can be discovered. Further discussion about such outcomes is presented in \autoref{sec:classification}.

\begingroup
\medmuskip=0mu
\thickmuskip=0mu
\begin{table}[!ht]
\centering
\scriptsize
\caption{The average weight denotes the relevance (in percentage) of certain features in a corresponding classification setting. Since the alphanumeric sequences are used as the input to compute the rest of the features, they do not have any weights. }
\label{tab:feature_w}
%\rowcolors{2}{gray!25}{white}

\begin{tabular}{>{\bfseries}cp{1.2in}cc}
\toprule
\textbf{Feature Set} & \textbf{Notation} & \textbf{\specialcell{Avg. Weight\\~~~(Binary)}}  & \textbf{\specialcell{Avg. Weight\\~(Multiclass)}}\\
\midrule
\multirow{5}{*}{Statistical Attributes}& $L-HEX$  & 1.38 & 0\\
& $L-LEN$  & 5.32  & 19.76 \\
& $L-DIG$ & 3.93 & 11.84  \\
& $L-DOT$  & N/A & 2.81\\
& $L-CON-MAX$  & 1.36  & 0.69\\
& $L-VOW-MAX$  & 0.32 & 0.18\\
& $L-W2$ & 0.45 & 1.01\\
& $L-W3$ & 1.59 & 0.07\\ %words longer than 3 chars in txt1
 \midrule
 \multirow{8}{*}{Ratios} & $R-CON-VOW$& 1.47 & 0.61\\
&$R-Dom-3G$   & 1.57 & 1.53\\   %
&$R-Dom-4G$   & 15.23 & 2.18  \\  %
&$R-Dom-5G$    & 11.70 & 0.43\\   %
& $R-VOW-3G$  & 0.75  & 0.36\\
& $R-VOW-4G$ & 0.65 & 0.33\\
& $R-VOW-5G$  & 0.62  & 0.26\\
&$R-WS-LEN$  & 7.95 & 2.42\\
&$R-WD-LEN$ & 2.61 &  9.35\\
&$R-WDS-LEN$  & 6.20 & 7.40 \\
&$R-W2-LEN$  & 1.95 & 0.84\\
&$R-W2-LEN-D$  & 1.28 & 0.33\\
&$R-W3-LEN$  & 2.61 & 0.16 \\
&$R-W3-LEN-D$  & 2.54 & 0.16\\

\midrule
\multirow{10}{*}{Gibberish Probabilities}&$GIB-1-Dom$  & 5.12 & 0.48 \\
&$GIB-1-Dom-WS$   & 1.67 & 0.52\\% GIB1 the higher the better
&$GIB-1-Dom-D$   & 5.76  & 0.48\\
&$GIB-1-Dom-WDS$   & 0.82 & 0.48\\
&$GIB-1-Dom-W2$   & 0.46 & 0.46\\
&$GIB-1-Dom-W3$   & 0.97 & 0.07\\
&$GIB-2-Dom$   & 0.92 & 0.75\\    % gib2 the higher the worse
&$GIB-2-Dom-WS$   & 0.87& 0.80 \\
&$GIB-2-Dom-D$   & 1.09 & 0.69\\
&$GIB-2-Dom-WDS$   & 0.89 & 1.26\\
&$GIB-2-Dom-W2$   & 0.19 & 0.24\\
&$GIB-2-Dom-W3$   & 1.00 & 0.11\\
\midrule
\multirow{5}{*}{Entropy}&$E-Dom$   & 2.10 & 12.66\\
&$E-Dom-WS$  & 1.44 & 8.19\\
&$E-Dom-D$   & 2.27 & 4.76\\
&$E-Dom-WDS$   & 1.51 & 3.39\\
&$E-Dom-W2$  & 0.63 & 1.62\\
&$E-Dom-W3$  & 0.81 & 0.16 \\
\bottomrule
\end{tabular}
\end{table}
\endgroup

\end{document}